\documentclass{article}

\PassOptionsToPackage{numbers, sort&compress}{natbib}
\usepackage[preprint]{neurips_2026}

\usepackage[utf8]{inputenc} % allow utf-8 input
\usepackage[T1]{fontenc}    % use 8-bit T1 fonts
\usepackage{hyperref}       % hyperlinks
\usepackage{url}            % simple URL typesetting
\usepackage{booktabs}       % professional-quality tables
\usepackage{amsfonts}       % blackboard math symbols
\usepackage{nicefrac}       % compact symbols for 1/2, etc.
\usepackage{microtype}      % microtypography
\usepackage{xcolor}         % colors
\usepackage{amssymb}

\usepackage{amsmath}
\usepackage{graphicx}
\usepackage{wrapfig}

\newcommand{\RR}{\mathbb{R}}
\newcommand{\cL}{\mathcal{L}}

\newcommand{\Id}{\mathbb{I}}
\newcommand{\softmax}{\mathrm{softmax}}

\newcommand{\ee}{\mathrm{e}}
\newcommand{\diff}{\mathrm{d}}
\newcommand{\myvec}[1]{\mathbf{#1}} % vectors 
\newcommand{\mymat}[1]{#1} % matrices \mathsf{#1}

\DeclareMathOperator{\Tr}{Tr}

\title{Learning Dynamics of Chain-of-Thought State Tracking in a Solvable Transformer Model}

\author{%
  Niklas Forner,\textsuperscript{1} \hspace{1ex}
  Marcel K\"uhn,\textsuperscript{1, 2} \hspace{1ex}
  Matthias Thamm,\textsuperscript{1} \hspace{1ex}
  \hspace{-1ex} and \hspace{0.5ex}
  Bernd Rosenow\textsuperscript{1, 2} \\
  $^1$Institute for Theoretical Physics, University of Leipzig, D-04103 Leipzig \\
  $^2$ScaDS.AI Dresden/Leipzig, University of Leipzig, D-04105 Leipzig
}

\begin{document}

\maketitle

\begin{abstract}
    Chain-of-thought generation can turn a multi-step computation into a sequence of locally checkable state updates, but the training dynamics by which transformers acquire such updates remain poorly understood. 
    We study this question in a solvable setting: a simplified one-block transformer trained by supervised next-token prediction on state sequences generated by composing permutations. 
    The architecture separates fixed-lag action retrieval, learned by RoPE attention, from a specialized MLP logic module that applies the retrieved permutation to the current state. 
    Using a statistical-physics mean-field description, we derive closed dynamics for three order parameters measuring attention retrieval, teacher-matrix alignment, and off-target logic overlap. 
    These equations quantitatively match simulations for the order parameters and, combined with a logit-distribution approximation, qualitatively predict the sharp transition in final rollout accuracy. 
    The analysis reveals staged learning: the logic module first learns a mixed heuristic; attention then locks onto the relevant action, enabling efficient MLP alignment. 
    Together, these results provide a controlled mechanistic account of how attention-based retrieval and MLP-based logic co-develop during chain-of-thought state tracking.
 \end{abstract}

% ======================================================================

\section{Introduction}
\label{sec:introduction}

Transformers \cite{vaswani2017attention} are highly successful across tasks such as translation, text classification, and question answering \cite{yang2019xlnet, touvron2023llama, le2023bloom, Liu.2019a}.
In particular, chain-of-thought reasoning \cite{wei2022chain} has been a major recent advance:
by inserting intermediate tokens that support step-wise answer production, large language models (LLMs) substantially improve on complex tasks \cite{wei2022chain}. 
Alongside in-context learning \cite{brown2020language, von2023transformers, chan2022transformers, raventos2023pretraining} and grokking \cite{power2022grokking}, the sharp performance gain from chain-of-thought reasoning can be viewed as an emergent phenomenon.
% This success ?
This performance has motivated much work in explainable AI \cite{elhage-2021-circuits, neo2024interpreting, yun2020transformers, geva2021ffnkv, wang2022interpretability, olsson2022context, bereska2024mechanistic, sharkey2025open, chughtai2023toy}, since the internal processes of transformers and LLMs are highly complex.
To understand why these systems perform so well, researchers have analyzed them empirically and investigated simplified settings analytically.
Prior work \cite{yun2020transformers, neo2024interpreting, meng2022locating} suggests that the attention block and multi-layer perceptron (MLP) inside a transformer serve two core functions: attention is mainly responsible for semantic and positional structure, while the MLP largely handles information storage and processing \cite{yun2020transformers, neo2024interpreting, meng2022locating}.
Of course, these two mechanisms do not have completely separate roles; they depend on each other, and their learning dynamics are coupled during training.
This raises basic questions: how do attention and MLP influence one another during training, and when do they learn relative to each other?

In this work, we disentangle these two core parts of a transformer. 
We let the attention mechanism handle positional information, deliberately without semantic complexity, and let the subsequent MLP act as an information storage and logic module. As a specific problem, we employ a state-tracking task in which the model sequentially composes permutations.
This compositional task has been used before to study length generalization of chain-of-thought computation in regular transformers \cite{huang-2025-cotlength}. 
Due to its exact, discrete nature and its local state dependence on the prior tokens, this task is well suited for detailed analysis.
The sequential application of group elements is a problem that can be solved by transformers \cite{marchetti2026sequential, chughtai2023toy, huang-2025-cotlength}, and in fact by a one-block transformer.
Here, we study this problem in a student-teacher setup, through the lens of permutation composition, where an exact teacher is known.
By fixing selected parts of the architecture of a one-block transformer, we obtain a solvable model that separates positional retrieval from logical composition learning.
In doing so, we study a controlled mechanistic model of serial state tracking via intermediate tokens.
% \paragraph{Scope}
Our goal is not to model natural-language chain-of-thought reasoning in full-scale language models. 
We instead isolate the core of algorithmic chain-of-thought tasks: retrieving an operation from context and applying it to an evolving state. 
Fixed embeddings and values, together with the specialized MLP-style logic module, are deliberate simplifications that make the retrieval-logic coupling analytically tractable. The resulting model should therefore be read as a solvable mechanistic model of chain-of-thought state tracking, rather than as a complete model of standard transformer reasoning.

\paragraph{Main results and contributions.} 
(1) We study a specialized, solvable transformer model in which attention learns positional information and an MLP-style block learns a permutation-composition task.
(2) We describe the learning process with three order parameters and derive differential equations for their training dynamics.
(3) During training, the model first learns a mixed heuristic in the MLP block; later, attention learns the relevant positional retrieval while the MLP logic aligns with the teacher. 
(4) We model the steep increase in final rollout accuracy, giving a simple explanation for a rapid emergent change as a function of training time.

\section{Related work}
\label{sec:related-work}

\paragraph{Explainable AI.} 
Mechanistic interpretability aims to understand neural networks by reverse-engineering the algorithms they implement \cite{elhage-2021-circuits, neo2024interpreting, yun2020transformers, geva2021ffnkv, wang2022interpretability, olsson2022context, bereska2024mechanistic, sharkey2025open, chughtai2023toy}. 
Prior work has identified different roles for attention, which performs contextual mapping, and the MLP, which performs value mapping \cite{yun2020transformers}.
Some works consider attention-only transformers \cite{elhage-2021-circuits, wang2022interpretability}, while others investigate factual knowledge stored in MLP layers \cite{meng2022locating}.
\citet{geva2021ffnkv} show that feed-forward layers in transformer-based language models operate as key-value memories \cite{sukhbaatar2019augmenting}.
\citet{neo2024interpreting} analyze interactions between MLP and attention layers and find that some attention heads identify context cues relevant to next-token prediction and, in response, activate downstream neurons that support that prediction.

\paragraph{Analytical models for transformers and limiting cases.} 
A range of theoretical transformer models and limiting cases has been studied.
\citet{poc2024dynamical} employ a dynamical mean-field approximation for large self-attention networks by mapping attention to an asymmetric Hopfield network \cite{treves1988metastable}.
\citet{maloney2022solvable} study a solvable statistical model combining a generative data model and a random feature model, in order to derive neural scaling laws.
In-context learning is studied by \citet{lu2025asymptotic} using a simplified linear self-attention model.

\citet{yang-2024-words} analyze gradient-flow training of a shallow transformer on word co-occurrence recognition and identify a two-phase dynamic in which a linear MLP first aligns with task-relevant target signals while softmax attention remains nearly static; later, attention and MLP jointly increase the margin and drive the loss toward its minimum.    
\citet{chen-2026-condensation} similarly argue for staged transformer optimization in a linearized setting, showing that attention parameters first condense toward task-relevant directions before key--query matrices participate more actively and normalized weights collapse toward low rank.   
Another transition appears in induction-head formation, where training moves from a "lazy'' local $n$-gram strategy to a richer in-context mechanism; this again suggests that distinct transformer submodules may enter the learned computation at different points in training \cite{wang-2025-rich}. 
Our work is complementary: rather than analyzing co-occurrence, rank collapse, or induction heads in isolation, we study a chain-of-thought state-tracking task in which the model must maintain and update a latent state across steps.
We pair finite-width training runs with analytical predictions that expose a two-stage division of labor between attention-mediated action access and MLP-mediated logical update.

\paragraph{Empirical studies on attention and MLP dynamics.}
\citet{tian-2024-empiricaldynamics} study joint MLP/attention dynamics in multilayer transformers and report that MLP nonlinearity and residual structure shape how attention evolves during training, supporting the view that feed-forward and attention modules interact rather than learn independently.
\citet{chen-2025-distributional} provide an empirical parallel: in a controlled next-token prediction setting, feed-forward layers preferentially learn simple distributional associations such as bigrams, whereas attention layers are more closely tied to in-context reasoning.
This is consistent with a temporal and functional separation between early MLP learning and later attention-mediated computation.
\citet{hakimi-2025-mechinterp} find a progression from general-purpose components to specialized factual components at larger model scale, with FFNs remaining comparatively stable and attention heads showing higher turnover. Their results likewise suggest staged component specialization rather than uniform co-development.

\paragraph{Reasoning theory and chain of thought.} 
Recent progress in LLM reasoning has been facilitated by the observation that transformers often solve complex tasks better when they are allowed to proceed step by step \cite{nye-2021-intermediatesteps}.
Chain-of-thought reasoning can produce a qualitative jump in reasoning performance \cite{wei2022chain}.
Transformers are capable of in-context learning \cite{lampinen2022can}, and intermediate steps can also help models improve their outputs iteratively \cite{NEURIPS2022_639a9a17}. 
\citet{prystawski2023think} investigate how chain-of-thought reasoning improves output quality and show that intermediate steps are helpful when the training data consist of overlapping, strongly interacting local clusters of variables.
Although transformers can learn some reasoning tasks using far fewer layers than the number of reasoning steps \cite{liu2022transformers}, they cannot learn certain arithmetic tasks without scaling the depth super-polynomially with input size \cite{feng2023towards}.
\citet{nadgir2026does} model chain-of-thought reasoning as a tree process and identify a critical branching-degree threshold: below this threshold, reasoning hurts performance, while above it, an optimal tree depth minimizes error.
Interestingly, constant-depth transformers can learn serial problems such as permutation composition through chain of thought \cite{li2024serialcot, feng2023towards}.

\paragraph{Permutation models and permutation composition.}
Many tasks require compositions of operations, whether mathematical, algorithmic, or otherwise. 
The solvability of such sequential group-composition tasks depends on layer depth and architecture \cite{marchetti2026sequential}.
\citet{chughtai2023toy} investigate how small networks encode finite group operations.
\citet{liu2022transformers} show that some composition tasks may be implemented efficiently by the model using shortcuts.
This matters because a network may solve the training task while failing to learn the underlying rule, and hence fail to generalize.
In this work, we study a state-tracking task in which we can directly follow which state the network is tracking at every step, leaving little room for hidden shortcuts.

Prior work on chain-of-thought and permutation/group-composition tasks primarily addresses expressivity, learnability, length generalization, or reverse engineering of trained circuits. In contrast, our focus is the gradient dynamics by which a transformer-like architecture separates the computation into retrieval and state-update components. The main question is not whether the task can be solved, but how attention and the MLP-style logic module co-evolve during learning, and why final rollout accuracy undergoes a sharp transition.

\begin{figure}[!tb]
    \centering
    \includegraphics[width=1\linewidth]{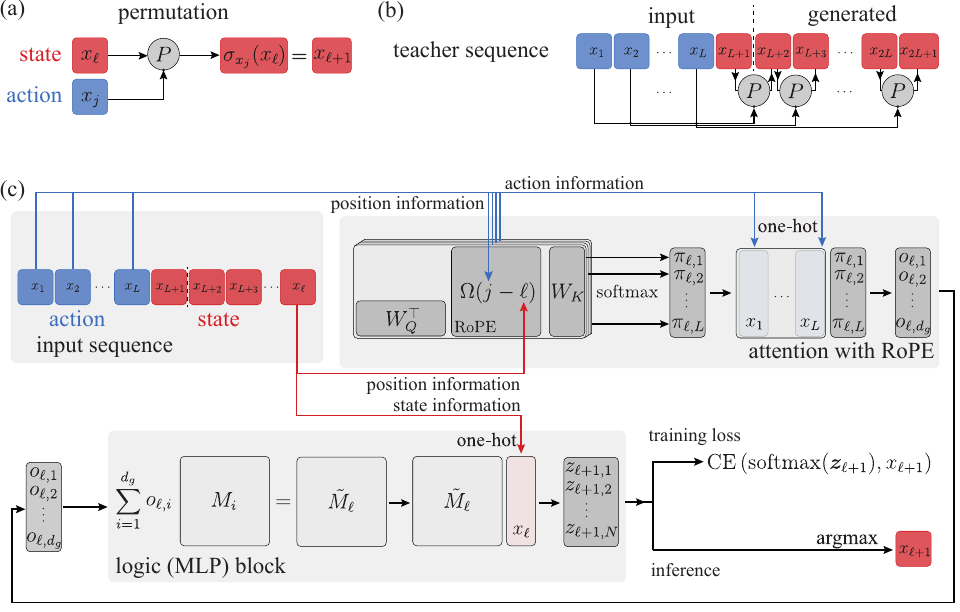}
    \caption{Problem setup and specialized one-block transformer model.
    (a) Action token $x_j$ selects permutation $\sigma_{x_j}$, which maps current state $x_\ell$ to next state $x_{\ell+1}$.
    (b) Teacher sequence from actions $x_1,\dots,x_L$ and initial state $x_{L+1}$; for $i=L+1,\dots,2L$, $x_{i+1}$ is obtained by applying $x_{i-L}$ to $x_i$.
    (c) Student prediction: RoPE attention retrieves the action at fixed lag $-L$, and the logic block applies the corresponding weighted sum of learnable matrices to the one-hot state, producing logits $z_{\ell+1}$.
    }   
  \label{fig:setup}
\end{figure}

\section{Problem and network setup}
\label{sec:problem-setup}

To isolate the model's ability to track latent states, we construct a synthetic environment that serves as a symbolic chain of thought.
It retains the core difficulty of serial reasoning: a sequence of intermediate steps must be carried out before the final answer is reached.
This yields a teacher with exact state tracking and a student trained to reproduce the resulting chain.

\subsection{Permutation-chain task}
Consider the first $N \ge 2$ positive integers $[N] := \{1, \dots, N \}$ and fix $d_g \leq N!$ permutations $\{ \sigma_{i} \}_{i = 1, \dots, d_g} \subseteq S_N$ acting on $[N]$.  This subset forms the action vocabulary for the task.
We encode a sequence of permutations, called \textit{actions}, and the resulting \textit{states} as a chain of tokens that the network must reproduce.
First, we draw $L \in \mathbb{N}$ actions (i.i.d.) from $\{ \sigma_{i} \}_{i = 1, \dots, d_g}$ and represent them by their corresponding indices $g_1, \dots, g_L \in \{ 1, \dotsc, d_g \}$. 
Second, we choose an initial state $s_0 \in [N]$.
Third, we recursively generate $L$ states by applying the actions in chain form:
\begin{equation}
    s_t = \sigma_{g_t}(s_{t-1}), \qquad t=1,\dots,L.
\label{eq:teacher_recursion}
\end{equation}
This trajectory is concatenated into a single token sequence of total length $2L+1$:
\begin{equation}
    x_1 = g_1, \ \dots\ ,\ x_L = g_L, 
    \quad 
    x_{L+1} = s_0,
    \quad 
    x_{L+2} = s_1, \ \dots \ , \ x_{2L+1} = s_L.
\label{eq:sequence-format}
\end{equation}
Given the actions and the initial state, the student is tasked with generating the $L$ subsequent states, ending with the final rollout token $s_L$. 
Thus, at inference, the prompt is the prefix $(g_1, \dots, g_L, s_0)$ and the output is the generated suffix $(s_1, \dots, s_L)$.
This chain can be generalized to the sequential application of group elements \cite{marchetti2026sequential}.
Here, we focus on permutations, since they admit an exact matrix representation.

\subsection{Network setup}
We consider a minimal one-block transformer in a student-teacher setting, in which a modified attention head retrieves the relevant action at a fixed lag and passes it to an explicit logic module.
This logic module performs the next-token prediction, thereby isolating retrieval from application.
A single-layer transformer can encode the solution \cite{huang-2025-cotlength}; we use this minimal depth to focus on the reasoning process through intermediate steps.
Figure \ref{fig:setup} summarizes the setup: the chosen permutations determine the teacher sequence, and the student predicts each next token by combining attention-based retrieval with the logic block.

\paragraph{The attention head}
We use a single attention head to retrieve the relevant action from the prefix. 
We choose Rotary Positional Encoding (RoPE) \cite{su2024roformer} and fix the values in order to focus on positional information.
Compared with absolute positional encodings, relative or rotary encodings have been found to perform better \cite{allenzhu-2025-physicslanguagemodel}.
We define
\begin{align}
    \myvec{q_i} &:= \mathrm{RoPE}_i \! \left( \mymat{W_Q} \, \myvec{u_i} \right) \in \RR^{d_h}
    \ , \quad 
    \myvec{k_j} := \mathrm{RoPE}_j \! \left( \mymat{W_K} \, \myvec{u_j} \right) \in \RR^{d_h}
    \ , \quad 
    \myvec{v_j} := \mathrm{Val}(x_j) \in \RR^{d_g} \ ,
\end{align}
where $d_g$ is the size of the action vocabulary, $d_h$ is the head dimension, and $\myvec{u_i}$ is the embedding of token $x_i$.
The values are defined as one-hot features for action tokens and as zero for state tokens, i.e., $\mathrm{Val}(g_j) = \myvec{\hat{e}_{g_j}} \in \RR^{d_g}$ for an action and $\mathrm{Val}(x_j) = \myvec{0}$ otherwise. 
To isolate positional information from semantic structure, we hold the embedding vectors fixed at $\myvec{u_i} = E(x_i) = 1 \in \RR^1$ for all $i = 1, \dotsc, 2L+1$, which reduces the key and query matrices to vectors.
Appendix \ref{app:mlp-construction} explains how this can be achieved with bundled embeddings and matrix-shaped keys and queries.
Causal attention entries and the attention-head output are given by
\begin{align}
    \pi_{ij} = \softmax_{j \le \min \{ i, L \}} \biggl( \frac{\myvec{q_i^{\top}} \myvec{k_j}}{\sqrt{d_h}} \biggr)
    , \qquad
    \myvec{o_i} := \sum_{j \le i} \pi_{ij} \, \myvec{v_j} \ .
\end{align}
This $d_g$-dimensional output assigns a weight to each possible action. 
Perfect learning corresponds to a one-hot output identifying the correct action among all $d_g$ possibilities.

\paragraph{Positional encoding}
We apply RoPE to queries and keys, i.e., there are no explicit learned positional embeddings.
Let the query/key head dimension be $d_h \in 2 \mathbb{N}$; then the encoding is realized by a matrix with $d_h / 2$ planar rotation blocks.
RoPE at sequence position $i$ is the linear map $\mathrm{RoPE}_i(\myvec{W}) := \mymat{\Omega(i)} \, \myvec{W}$, where $\myvec{W}$ is a query or key vector and $\mymat{\Omega(i)}$ is the matrix of block rotations.
For attention, the relevant quantities are the query-key dot products.
They obey the identity
\begin{align}
    \big( \mathrm{RoPE}_i (\myvec{W_Q}) \big)^\top \big( \mathrm{RoPE}_j (\myvec{W_K}) \big) = \myvec{W_Q^\top} \, \mymat{\Omega(j-i)} \myvec{W_K} \ .
\end{align}
The attention matrix therefore depends on the relative offset $j-i$ in the token sequence, which makes RoPE suitable for a fixed-lag retrieval task.

\paragraph{The logic module}
The attention output is supplied to a logic module to produce a prediction of the next state.
The module can be constructed as a constrained instantiation of a GLU-style MLP; see Appendix \ref{app:mlp-construction} for the explicit construction.
It is convenient to represent the permutations by matrices. 
Applying a permutation can then be written as $\mymat{O(\sigma_{a})} \, \myvec{e_j} = \myvec{e_{\sigma_a (j)}}$ for $j \in [N], \ a \in [d_g]$, where $\{ \myvec{e_j} \}_{j=1}^N$ is the standard basis of $\RR^N$. 

We introduce the set $\{ \mymat{M_a} \}_{a=1}^{d_g}$ of learnable matrices $\mymat{M_a} \in \RR^{N \times N}$. 
With attention outputs $\myvec{o_{i}} \in \RR^{d_g}$, define the linear map
\begin{equation}
    \mymat{\tilde{M}_i} := \sum_{a=1}^{d_g} (o_{i})_a \, \mymat{M_a} \ \in \RR^{N\times N} \ .
\end{equation} 
When the student has learned, this map approximates the correct permutation matrix.
Before convergence, it is a weighted sum of the learnable matrices associated with the action vocabulary.
Actions identified by the attention head as relevant receive larger weights than irrelevant ones.
The next-token computation is performed by matrix multiplication and uses a one-hot encoding of the states.
We represent the current state $x_{i} = s_{i-L-1}$ as an $N$-dimensional one-hot vector $\myvec{e_{s_{i-L-1}}} \in \RR^N$ and define $\myvec{z_{i+1}} := \mymat{\tilde{M}_{i}} \, \myvec{e_{s_{i-L-1}}} \in \RR^N$.
The prediction is obtained by choosing the index of the maximal entry.
We denote the prediction by $\hat{s}_{i-L}$.

This setup preserves the core functionalities of a standard transformer block consisting of an attention head followed by an MLP.
During training, the model sees every intermediate token.
At inference time, it is given only the prefix $(g_1, \dots, g_L, s_0)$ and must generate $s_1, \dots, s_L$; the final answer is the last generated token $s_L$. 
In this way, the student must carry out a chain-of-thought process to arrive at the last state.
Under teacher forcing, the previous state token is locally available and need not be retrieved by a separate transport head.
A more general model could include a second attention head tasked with copying this previous state token.
In the teacher-forced regime, the current state is already available at the predictor token, so a single nontrivial fixed-lag retrieval for $g_t$ suffices.

\subsection{Teacher model}
We specify a teacher within the same architecture that implements the correct dynamics $s_t = \sigma_{g_t}(s_{t-1})$.
To distinguish teacher from student parameters $\theta$, we denote teacher parameters by $\theta^\star$.

\paragraph{Attention.} 
We want the action head to attend from predictor positions $j + L$ to action positions $j$, i.e., to the relative offset $-L$. 
This corresponds to $\pi_{j + L, j}^{\star} = 1$ for $j= 1, \dots, L$ and zero everywhere else, and can be realized arbitrarily well by an appropriate choice of keys and queries.
Then, the attention-head output is $\myvec{o^{\star}_{i}} = \myvec{v_{i-L}} \equiv \myvec{\hat{e}_{g_{i-L}}}$, using the same value map as for the student.

\paragraph{Next tokens.}
In the teacher logic module, the matrices are set to the exact permutation matrices, $\mymat{M_a^\star} := \mymat{O(\sigma_a)} \in \RR^{N\times N}$ for $a=1,\dots,d_g$.
Consequently, the per-step map becomes $\mymat{\tilde{M}_{i}^{\star}} = \mymat{M_{g_{i-L}}^\star}$ and hence equals the exact permutation matrix $\mymat{M_{g_{i-L}}^\star} =\mymat{O(\sigma_{g_{i-L}})}$.
Therefore, the teacher specifies the next token $\myvec{p_{i+1}^\star} = \mymat{O(\sigma_{g_{i-L}})} \, \myvec{e_{s_{i-L-1}}} = \myvec{e_{s_{i-L}}}$ as the analog of the student logit, the target vector. 
The position of the maximal entry is the correct next state $s_{i-L}$.

\subsection{Training objective}
We train the model autoregressively with teacher forcing on full sequences \eqref{eq:sequence-format}, but compute the loss only on the suffix state tokens $s_1, \dots, s_L$ ($s_1 = x_{(L+1)+1}, \dots, s_{L} = x_{2L+1}$).
We use the cross-entropy loss 
\begin{equation}
    \cL(\theta)
    = \Biggl\langle \frac{1}{L} \sum_{i=L+1}^{2L} - \log P\!\left( x_{i+1} \mid x_1, \dots, x_{i} \right) \Biggr\rangle_{x_1, \dots, x_{2L+1}}
\label{eq:loss}
\end{equation}
with $\theta$ denoting the network parameters.
The current state token $x_{j+L} = s_{j-1}$ is used to predict the next token.
To do so, the model needs the corresponding action token $x_{j} = g_{j}$ to apply inside the logic module.
The required relative offset is therefore $j - (j+L) = - L$, a single fixed lag independent of the current position.
The model performs this next-token prediction at all $L$ state positions. 
For the hyperparameters used in simulations, the reader is referred to Table \ref{tab:config} in the appendix.
Details on dataset generation are given in Appendix \ref{app:setup}.

\section{Macroscopic order parameters and mean-field formulation}
\label{sec:order-params}

\subsection{Order parameters} 

\paragraph{Attention order parameter} 
At prediction positions $i = L+1, \dots, 2L$, the fixed-lag structure identifies a unique correct action position, namely $i-L$. 
We therefore define
\begin{align}
    A := \frac{1}{L} \sum_{i = L + 1}^{2L} \pi_{i, i-L} \ .
\end{align}
This order parameter measures how much attention mass the head assigns, on average, to the correct action across all prediction steps.
Because the attention entries are normalized, $A = 1$ corresponds to perfect attention: at each step, the model retrieves the correct action.
When $A \approx 0$, the model actively avoids the correct action.
Up to fluctuations, $A = 1/L$ corresponds to random guessing over the $L$ candidate action positions.

\paragraph{Logic order parameters}
Recall the student and teacher matrices $\{ \mymat{M_a} \}_{a = 1}^{d_g}$ and $\{ \mymat{M_a^\star} \}_{a = 1}^{d_g}$. 
Let
\begin{subequations}
    \begin{align}
        R &:= \frac{1}{d_g} \sum_{a = 1}^{d_g} \frac{1}{N} \Tr \Bigl[ \bigl( \mymat{M^{\star}_{a}} \bigr)^{\top} \mymat{M_a} \Bigr] \ ,
        \\
        S &:= \frac{1}{d_g (d_g - 1)} \sum_{a = 1}^{d_g} \sum_{\substack{b = 1 \\ b \neq a}}^{d_g} \frac{1}{N} \Tr \Bigl[ ( \mymat{M^{\star}_{a}} )^{\top} \mymat{M_{b}} \Bigr] \ .
    \end{align}
\end{subequations}
$R$ measures the alignment between the student's learned matrices and the corresponding teacher permutation matrices.
$S$ quantifies the off-target overlap between learned matrices and the other teacher permutation matrices.

% Figure for Simulations
\begin{figure}[!tb]
    \centering
    \includegraphics[width=1\linewidth]{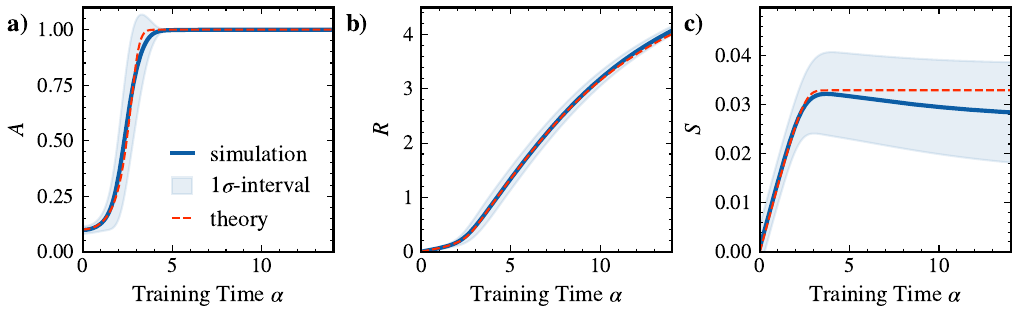} %Dynamics-constant-prefactor-approx.png}
    \caption{Dynamics of the order parameters $A$, $R$, and $S$ in panels a), b), and c), respectively. 
    Simulations are averaged over 100 seeds, and $\sigma$ denotes the standard deviation.
    The theoretical predictions agree very well with the simulations (note different scale in panel c)). 
    }
    \label{fig:dynamics-constant-prefactor-approx}
\end{figure}

\subsection{Mean-field loss and logits}
Denoting the model-output pre-activations as $\myvec{z_{i+1}}$ at position $i$, the cross-entropy loss is given by $\mathcal{L} = \frac{1}{L} \sum_{i = L+1}^{2L} [- \myvec{p^{\star}_{i+1}} \cdot \, \myvec{z_{i+1}} + \ln (\sum_{r = 1}^{N} \ee^{z_{i+1, r}})]$.
Here, the teacher specification $\myvec{p^{\star}_{i+1}}$ singles out the correct logit.
To obtain an ensemble description, we replace the seed-specific loss by an ensemble-averaged one.
We make a mean-field ansatz and replace the correct and other logits by their expectation values with respect to states and actions.
The resulting mean-field loss is
\begin{align}
    \ell_{\mathrm{CE}} &:= \frac{1}{L} \sum_{i = L+1}^{2L} - \langle z_{i+1, k} \rangle + \ln \biggl( \ee^{\langle z_{i+1, k} \rangle} + (N-1) \ee^{\langle z_{i+1, \overline{k}} \rangle} \biggr) \ ,
\end{align} 
where the correct logit is denoted by $z_{i+1,k} = \myvec{p_{i+1}^{\star}} \cdot \myvec{z_{i+1}}$.
By symmetry, the other logits behave identically on average so that $z_{i+1, \overline{k}}$ describes any logit but the correct $z_{i+1,k}$.
In this way, the loss can be expressed in terms of the order parameters together with an initialization-dependent constant.
The mean-field loss describes both the single-seed loss and the seed-averaged loss very well; see Fig.\,\ref{fig:mean-correct-and-other-logits-and-loss} in the appendix.

For the correct next-token logit $z_{i+1, k}$, the teacher logit vector $\myvec{p_{i+1}^{\star}}$ is one-hot, with unity at entry $k$. 
It can be expressed using the exact teacher construction.
The expectation value is taken over the local action and state.
In principle, it can be reduced to an average over the actions and the initial state, since permutation composition preserves the relevant statistics.
These variables are statistically independent, and a uniform distribution suffices to describe the random drawing of input tokens.
In the ensemble, no step in the state sequence is preferred, so $\pi_{i, i-L} = A$ describes the relevant attention entries well.
This behavior can also be observed directly during training; for details, see Fig.\,\ref{fig:attention-entries} in the appendix.
Ultimately, this yields the mean correct logit:
\begin{align}\label{eq:mean-correct-logit}
    \langle z_{i+1, k} \rangle &= \biggl( A + \frac{1 - A}{d_g} \biggr) R + \frac{d_g - 1}{d_g} \, (1 - A) S \ .
\end{align}
The dependence on the sequence position $i$ disappears after averaging.

The mean other logits can be obtained from the average total logit, independently of which state label is correct.
Because the sum of the logits remains constant under cross-entropy training without weight decay, the correct logit is closely tied to the remaining ones.
Explicitly, we find
\begin{align}\label{eq:mean-other-logits}
    \langle z_{i+1, \overline{k}} \rangle  &= - \frac{\langle z_{i+1, k} \rangle}{N - 1} + \frac{1}{d_g N (N-1)} \sum_{a = 1}^{d_g} \sum_{r, s = 1}^{N} \bigl[ \mymat{M_{a}} \bigr]_{rs} \ ,
\end{align}
where the sum of learnable matrices depends on the initialization and remains constant throughout training.
We expect the mean correct logit to grow during training.
When it does, the remaining logits decrease, so the network progressively suppresses all incorrect classes.

\subsection{Training dynamics}
Autoregressive training induces discrete updates of the microscopic network parameters in the negative gradient direction of the exact loss.
Under the mean-field ansatz, the update for a representative learnable matrix entry is
\begin{align}
    [M_{a}]_{rs} \mapsto [M_{a}]_{rs} + \Delta [M_{a}]_{rs} 
    \ , \qquad 
    \Delta [M_{a}]_{rs} = - \eta \, \frac{\diff \ell_{\mathrm{CE}}}{\diff [M_{a}]_{rs}} \ , 
\end{align} 
where $\eta$ is the learning rate and $\ell_{\mathrm{CE}}$ is the mean-field loss as a function of all network parameters.
We define the \emph{training time} $\alpha$ by $\alpha := \text{\# epochs} / (d_g N)$.
Then one epoch corresponds to $\Delta \alpha := 1 / (d_g N)$, a small change in training time.
The mean-field loss allows us to translate these microscopic updates into dynamics for the macroscopic order parameters.
To that end, we assume that the attention mass on the correct action is exchangeable across sequence positions and described by $A$.
We average teacher-teacher matrix overlaps over seeds, and we close the attention-gradient factor as
 $d_g N [\vert \nabla_{\myvec{W_Q}} A \vert^2 + \vert\nabla_{\myvec{W_K}} A \vert^2] = L c_{\omega} A (1 - A)$ with 
\begin{align}\label{eq:c-omega-prefactor}
    c_{\omega} &:= \frac{2 d_g N}{L (L - 1)} \, \sigma^2 \cdot \frac{2}{d_h} \sum_{n = 1}^{d_h / 2} \Biggl[ 1 - \frac{1}{L^2} \biggl( \frac{\sin(\omega_{n} L / 2)}{\sin(\omega_{n} / 2)} \biggr)^2 \Biggr]^2 
\end{align}
determined by the initialization.
Here, $\sigma^2$ is the variance of the key/query initialization and $\{ \omega_{n} \}_{n = 1}^{d_h / 2}$ are the RoPE frequencies.
In our case, the default \texttt{PyTorch} \texttt{nn.Linear} initialization gives $\sigma^2 = \nicefrac{1}{3}$.
%For sufficiently small learning rate, we can neglect higher-order terms in $\eta$ in $\Delta A$.
Consequently, $\Delta A$, $\Delta R$, and $\Delta S$ become functions of the order parameters themselves and the network hyperparameters.
We then take the thermodynamic limit $d_g, N, L, d_h \to \infty$ while keeping $d_g / N$, $N / L$, and $L / d_h$ finite.
In this limit, $\Delta \alpha \to 0$, which allows us to pass from difference quotients to continuous derivatives, e.g., from $\Delta S / \Delta \alpha$ to $\diff S / \diff \alpha$.
This step is justified when the difference quotient converges uniformly on compact time intervals \cite{HairerNorsettWanner1993}, which can be ensured in the gradient-flow limit.
Here, this is consistent with the finite-ratio assumptions on the hyperparameters.
Under this mean-field closure, the macroscopic order parameters satisfy
\begin{subequations}\label{eq:differential-equations}
    \begin{align}
        \frac{\diff A}{\diff \alpha} &= \eta \, \frac{1}{1 + \frac{1}{N - 1} \, \ee^{m(A, R, S)}} \frac{N}{N - 1} \frac{d_g - 1}{d_g} \, c_{\omega} \cdot L A (1 - A) (R - S) \ , \label{Eq:diffeq_A}
        \\
        \frac{\diff R}{\diff \alpha} &= \eta \, \frac{1}{1 + \frac{1}{N - 1} \, \ee^{m(A, R, S)}} \cdot \biggl( A + \frac{1 - A}{d_g} \biggr) \ , \label{Eq:diffeq_R}
        \\
        \frac{\diff S}{\diff \alpha} &= \eta \, \frac{1}{1 + \frac{1}{N - 1} \, \ee^{m(A, R, S)}} \cdot \frac{1 - A}{d_g} \ , \label{Eq:diffeq_S}
    \end{align}
\end{subequations}
where $m(A, R, S)$ is the logit margin: $m(A, R, S) = \frac{1}{L} \sum_{i = L+1}^{2L} (\langle z_{i+1, k} \rangle - \langle z_{i+1, \overline{k}} \rangle)$, 
which can be expressed through Eqs.\,\eqref{eq:mean-correct-logit} and \eqref{eq:mean-other-logits}.
In Fig.\,\ref{fig:dynamics-constant-prefactor-approx}, we compare empirical results with numerical solutions of these differential equations.
The initial conditions are $A(0) = 1/L$ and $R(0) = S(0) = 0$.
The agreement is very good.
We see a clear transition in the attention mechanism: at short times $\alpha \lesssim 1$, $A$ remains near its initial value $A(0)$; later, it rises rapidly to saturation.
In the same time regimes, $R$ first increases slowly and then quickly aligns with the teacher once the attention head has started learning.
By contrast, the overlap with the non-matching teacher matrices, $S$, increases relatively quickly at first and saturates once attention is nearly perfect.

\paragraph{Interpretation}
During early training, the model learns a heuristic in the logic module: the student matrices acquire mixed directions.
The attention output weights the actions in the sequence nearly uniformly, so the logic module cannot yet apply the correct action to the current state.
Instead, the next state is predicted by applying a mixture of actions, and both the correct overlaps $R$ and the off-target overlaps $S$ increase.
This changes once attention begins to distinguish the correct action from the others.
At that point, the learnable matrices begin to align strongly with the teacher matrices, which in turn reinforces correct action selection in the attention block.
This MLP-first regime is consistent with gradient-flow analyses of transformer training, in which task-relevant linear or MLP directions emerge before attention has appreciably specialized \cite{yang-2024-words}. The resulting transition resembles later-stage attention specialization, although here it is tied specifically to selecting the correct state-update rule rather than to co-occurrence recognition or induction-head formation \cite{chen-2026-condensation, wang-2025-rich}.

\section{Final rollout accuracy}
\label{sec:rollout-accuracy}
Ultimately, the model should correctly predict the final token in the sequence, once the intermediate steps are learned well enough; that is, it should carry out a chain of thought without making errors along the way.
In this section, we model the accuracy of the final token.
At each step in the sequence, the model may predict either the correct next state or an incorrect one.
Accordingly, the probability of a correct prediction obeys the recursion
\begin{align}\label{eq:recursion-probabilities}
    P(\hat{s}_{t+1} = s_{t+1}) 
    &= P(\hat{s}_{t} = s_{t}) \cdot P(\hat{s}_{t+1} = s_{t+1} | \hat{s}_{t} = s_{t}) \nonumber \\
    &\quad + \bigl[ 1 - P(\hat{s}_{t} = s_{t}) \bigr] \cdot P(\hat{s}_{t+1} = s_{t+1} | \hat{s}_{t} \neq s_{t}) \ .
\end{align}

\begin{wrapfigure}{r}{0.48\columnwidth}
    \centering
    \includegraphics[width=\linewidth]{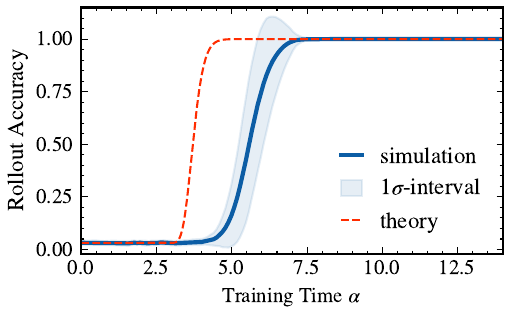}
    \caption{Final rollout accuracy, averaged over 100 model seeds; $\sigma$ denotes the standard deviation. 
    The theoretical curve uses $\mu_k$ and $\mu_{\overline{k}}$ from the order-parameter solutions of Eqs.\,\eqref{eq:differential-equations} and constant variances $\sigma_k^2=\sigma_{\overline{k}}^2\approx 0.044$ from initialization. 
    The predicted rise occurs slightly too early because the empirical variances increase during training; see Appendix \ref{app:rollout-accuracy-derivation}.}
    \label{fig:rollout-accuracy-constant-prefactor-approx}
\end{wrapfigure}

Since the initial token is given, we have $P(\hat{s}_{0} = s_{0}) = 1$. 
In principle, the model can recover the correct token even when acting on an incorrect current state, although this effect becomes less important as training progresses.

Both conditional probabilities in Eq.\,\eqref{eq:recursion-probabilities} depend on the logit distributions.
Empirically, the logits are approximately normally distributed at initialization and during the early stages of training.
Assuming that the correct logit follows a Gaussian distribution $\mathcal{N}(\mu_k, \sigma^2_k)$ and that each incorrect logit follows $\mathcal{N}(\mu_{\overline{k}}, \sigma^2_{\overline{k}})$, we can model these conditional probabilities.
For the next-token accuracy, we obtain $P(\hat{s}_{t+1} = s_{t+1} | \hat{s}_{t} = s_{t}) = \rho$ for any $t \in \{ 0, \dotsc, L-1 \}$, with
\begin{align}
    \rho
    &= \int_{-\infty}^{+\infty} \diff u \, \frac{\ee^{- u^2 / 2}}{\sqrt{2 \pi}} \, \Phi \! \biggl( \frac{\sigma_{k}}{\sigma_{\overline{k}}} \, u + \frac{\mu_{k} - \mu_{\overline{k}}}{\sigma_{\overline{k}}} \biggr)^{N-1} \ .
\end{align}
Furthermore, if the incorrect logits are symmetric and have no preference among themselves, the recovery term $P(\hat{s}_{t+1} = s_{t+1} \mid \hat{s}_{t} \neq s_{t})$ can be written as $(1 - \rho) / (N - 1)$.
Under this approximation, the final rollout accuracy becomes
\begin{align}\label{eq:rollout-accuracy}
    P(\hat{s}_{L} = s_{L})
    &\approx \frac{N - 1}{N} \biggl( \frac{N \rho - 1}{N - 1} \biggr)^{L} + \frac{1}{N} \ .
\end{align}
This expression is approximate for two reasons.
First, training introduces correlations between the correct and incorrect logits, whereas the derivation assumes independence.
Second, the logit distributions deviate from the Gaussian approximation as training proceeds.
Even so, the formula captures the qualitative form of the transition even for constant variances $\sigma^2_k, \sigma^2_{\overline{k}}$; see Fig.\,\ref{fig:rollout-accuracy-constant-prefactor-approx}.
The rollout accuracy remains near its initial plateau of $1/N$, then rises sharply and saturates at unity.
The reason for this sharp transition is clear from Eq.\,\eqref{eq:rollout-accuracy}:
the next-token accuracy enters to the $L$th power, so even a gradual increase in one-step accuracy is converted into a sudden jump in the final rollout accuracy.

\section{Discussion} 
\label{sec:discussion-and-limitations}

\paragraph{Summary}
This work studies a simplified transformer model for a compositional permutation task from a statistical-physics viewpoint. 
We introduce macroscopic order parameters to describe the learning dynamics of the attention head and the logic module (a specialized MLP).
We find that the model first learns a heuristic, before the attention mechanism aligns with the teacher and effective learning takes place.
We also derive a description of the accuracy of the final token in the chain of states:
a justified approximation reproduces the empirically observed sudden increase qualitatively and shows that this emergent phenomenon, as a function of training time, arises from the compositional structure of the task.
Taken together, these results shed light on the roles of the core components of a transformer in a controlled state-tracking problem.

\paragraph{Limitations}
The architecture studied here is deliberately constrained and therefore does not represent a standard one-block transformer, let alone a full-scale transformer.
Even so, it is plausible that similar learning phases may also occur in less specialized architectures on related tasks.
Our analytical results rely on a mean-field description of the loss, which introduces approximations relative to the exact training dynamics.
These approximations may behave differently in other models or on other tasks.
More work is needed to determine how far the present picture extends to standard transformer architectures, how robust the staged learning dynamics are, and how they are affected by reinforcement learning after pretraining.

% ======================================================================
% Acknowledgments 

\begin{ack}
% Use unnumbered first level headings for the acknowledgments. 
% All acknowledgments go at the end of the paper before the list of references. Moreover, you are required to declare funding (financial activities supporting the submitted work) and competing interests (related financial activities outside the submitted work).
% More information about this disclosure can be found at: \url{https://neurips.cc/Conferences/2026/PaperInformation/FundingDisclosure}.

% Do {\bf not} include this section in the anonymized submission, only in the final paper. You can use the \texttt{ack} environment provided in the style file to automatically hide this section in the anonymized submission.

M.K. acknowledges financial support by the Federal Ministry of Research, Technology and Space of Germany and by the Sächsische Staatsministerium für Wissenschaft, Kultur und Tourismus in the programme Center of Excellence for AI-research ``Center for Scalable Data Analytics and Artificial Intelligence Dresden/Leipzig'', project identification number: ScaDS.AI.
The work of B.R.~is supported in part by the Simons Foundation Grant No.~12574. 
% in the Simons Collaboration on the Physics of Learning and Neural Computation
\end{ack}

% by the excellence initiative ScaDS.AI funded by the BMFTR.

% ======================================================================

% References

% \bibliographystyle{plainnat}
\bibliographystyle{unsrtnat}
\bibliography{neurips_references}

% ======================================================================

\newpage
\appendix

\section{Hyperparameters}
\label{app:hyperparameters}

\begin{table}[h]
  \caption{Hyperparameters for the training simulations.}
  \label{tab:config}
  \centering
  \begin{tabular}{lll}
    \toprule
    Variable & Value & Description \\
    \midrule
    $N$ & $32$ & Magnitude of permutation set $\{1, \dots, N\}$ \\
    $L$ & $10$ & Sequence steps (actions $g_1, \dots, g_L$; predicted states $s_1, \dots, s_L$) \\
    $d_g$ & $32$ & Vocabulary size of permutations \\
    $d_{\text{model}}$ & $1$ & Embedding dimension \\
    $d_h$ & $128$ & RoPE encoding dimension (key \& query dimension) \\
    $n_{\text{train\_seq}}$ & $32{,}768$ & Number of training examples \\
    $n_{\text{test\_seq}}$ & $256$ & Number of test examples \\
    $\text{batch\_size}$ & $32{,}768$ & Batch size for gradient descent (full-batch if equal to $n_{\text{train\_seq}}$) \\
    $\text{lr}$ & $5 \times 10^{-1}$ & Learning rate \\
    $\text{weight\_decay}$ & $0$ & Weight decay (Adam optimizer) \\
    $\text{rope\_theta}$ & $10{,}000$ & RoPE frequency base (power-law distributed) \\
    $\text{train\_epochs}$ & $20{,}000$ & Number of training epochs \\
    $\text{eval\_every}$ & $50$ & Evaluation frequency (in epochs) \\
    $\text{seeds}$ & $100$ & Number of seeds (one model per seed) \\
    $\text{optimizer}$ & \texttt{sgd} & Optimizer type (\texttt{sgd} or \texttt{adamw}) \\
    % $\text{logit\_scale}$ & \texttt{False} & Learnable logit scaling parameter $a_u$ \\
    $\text{fix\_embedding}$ & \texttt{True} & Restrict embedding vectors to unity \\
    % $\text{with\_softmax}$ & \texttt{True} & Use of softmax in attention head output \\
    \bottomrule
  \end{tabular}
\end{table}
Table \ref{tab:config} lists all hyperparameters used for the chief simulations. 
The RoPE frequencies are power-law distributed and set as $\omega_{n} := 2 \pi \, \theta_{\mathrm{RoPE}}^{- 2 (n-1) / d_h}$ with $\theta_{\mathrm{RoPE}}$ (\verb|rope_theta|) as the base and $n = 1, \dots, d_h / 2$.

\section{Architecture Setup}
\label{app:setup}
% Explain setup in pytorch.
For the model training, we use \verb|PyTorch|.
For the sake of reproducibility, we sketch the numeric setup as follows:
\begin{itemize}
    \item Set seed with \verb|numpy|.
    \item Set RoPE frequencies.
    \item Draw permutation vocabulary ($d_g$) using \verb|rng.permutation(N)| with the default \verb|numpy| random number generator of the seed, then create the permutation matrices.
    \item Generate dataset (sequences): 
    \begin{itemize}
        \item Draw $L$ actions from the permutations and one initial index in $[N]$.
        \item Build the teacher token sequence by sequential application of the permutations on the current state.  
    \end{itemize}
    \item Define class with \verb|pytorch| using \verb|nn.Module|.
    \item Initialize two linear layers (keys and queries) and a parameter tensor for the learnable matrices uniformly.
    \item Forward function:
    \begin{itemize}
        \item Embed the token, compute key and query, apply RoPE.
        \item Compute attention matrix (causal and restricted to actions).
        \item Compute attention output with Values.
        \item Compute logic module output with learnable matrices and current token.
    \end{itemize}
    \item Train the model:
    \begin{itemize}
        \item Use train and test data loader for the dataset.
        \item Set optimizer (here \verb|SGD| with batch size equal to train set size)
        \item For each epoch, predict the tokens, compute the loss, apply the optimizer, backwards propagation, and optimizer step.
        \item Evaluate every few epochs the train/test accuracies, the order parameters, and other computable quantities.
    \end{itemize}
\end{itemize}

LLMs were used to write the code for training and analysis.
It was subject to scrutiny before execution, and the code performs the intended purposes.
The code is available in the supplementary material.
To train the model, execute the Python script with command arguments \verb|--seed <int> --dir "str"| for one seed, the output is saved as a numpy file in the given directory.
The hyperparameters are set within the script's code.
To inspect the data, compute the theoretical predictions, and generate the figures displayed in this work, use the Python Notebook.

For training, several GPUs and a cluster were used.
A T4 GPU takes 30 minutes to train one model (one seed), an NVIDIA Geforce RTX 4060 around 20 minutes.
We used seeds $0$ to $99$.

Many hyperparameter combinations were tested, hence more computational time was occupied than strictly necessary for the results shown here.
The model learns robustly for changes in the hyperparameters $N, d_g, L, d_h$ as well as the learning rate, RoPE frequency base and number of training examples.

Variations in $d_g, N$ (between $8$ and $32$) or $L$ (between $5$ and $20$) or $d_h$ (between $8$ and $128$) also produce fully learned models.
Fewer frequencies ($d_h$) limit the learnability.
Linearly distributed frequencies yield a similar learning behavior as the power-law distributed ones used here. 
Appendix \ref{app:robustness-hyperparameters} shows figures for another set of hyperparameters than used in the main text.

\section{Realizing the logic module as a constrained MLP}
\label{app:mlp-construction}

The logic module used in the main text differs from the feed-forward block of a standard transformer because the module output is written directly as a weighted sum of learnable matrices. 
Nevertheless, this computation can be implemented by a GLU-style MLP once the MLP input is arranged so that the first $d_g$ coordinates contain the action information $\myvec{o}\in\RR^{d_g}$ coming from attention and the last $N$ coordinates contain the current state $\myvec{s}\in\RR^N$. 
The state vector is one-hot, whereas $\myvec{o}$ may contain soft attention outputs.

Ignoring sequence indices, the desired logic computation is
\begin{align}
    \myvec{z}
    &:=
    \sum_{a=1}^{d_g} o_a \mymat{M_a} \myvec{s}
    =
    \bigl(\mymat{M_1}, \dots, \mymat{M_{d_g}}\bigr)
    \begin{pmatrix}
        o_1 \myvec{s} \\
        \vdots \\
        o_{d_g} \myvec{s}
    \end{pmatrix} \ ,
    \label{eq:logic-module-as-block-linear-map}
\end{align}
where $\mymat{M_a}\in\RR^{N\times N}$, $a = 1, \dots, d_g$, are the learnable action matrices. 
We now show that the vector in the second factor can be produced by fixed up- and gate-projections.

Consider a GLU-style MLP without biases,
\begin{align}
    \myvec{z}_{\mathrm{MLP}}
    &:=
    \mymat{W_d}
    \left[
        \mymat{W_u}
        \begin{pmatrix}
            \myvec{o}\\
            \myvec{s}
        \end{pmatrix}
        \odot
        \phi\!\left(
        \mymat{W_g}
        \begin{pmatrix}
            \myvec{o}\\
            \myvec{s}
        \end{pmatrix}
        \right)
    \right] \ .
    \label{eq:constrained-glu-definition}
\end{align}
We take the activation to be the identity, $\phi(x)=x$. 
The up-projection is fixed to copy the state vector into $d_g$ blocks,
\begin{align}
    \mymat{W_u}
    :=
    \begin{pmatrix}
        \mymat{0}_{N\times d_g} & \mymat{I}_{N} \\
        \vdots & \vdots \\
        \mymat{0}_{N\times d_g} & \mymat{I}_{N}
    \end{pmatrix}
    \in \RR^{d_g N \times (d_g+N)} \ , \qquad
    \mymat{W_u}
    \begin{pmatrix}
        \myvec{o}\\
        \myvec{s}
    \end{pmatrix}
    =
    \begin{pmatrix}
        \myvec{s}\\
        \vdots\\
        \myvec{s}
    \end{pmatrix} \ .
    \label{eq:fixed-up-projection}
\end{align}
The gate projection is fixed to copy each action coordinate over one state-sized block. 
Let $\mymat{T_a}\in\RR^{N\times d_g}$ denote the matrix whose $a$-th column is all ones and whose other entries are zero. 
Let $\myvec{1_N} \in \RR^N$ denote the all-ones vector. 
Set
\begin{align}
    \mymat{W_g}
    :=
    \begin{pmatrix}
        \mymat{T_1} & \mymat{0}_{N\times N}\\
        \vdots & \vdots\\
        \mymat{T_{d_g}} & \mymat{0}_{N\times N}
    \end{pmatrix}
    \in \RR^{d_g N \times (d_g+N)} \ , \qquad
    \mymat{W_g}
    \begin{pmatrix}
        \myvec{o}\\
        \myvec{s}
    \end{pmatrix}
    =
    \begin{pmatrix}
        o_1 \myvec{1}_N\\
        \vdots\\
        o_{d_g} \myvec{1}_N
    \end{pmatrix}.
    \label{eq:fixed-gate-projection}
\end{align}
Therefore, multiplying the two projections of Eqs.~\eqref{eq:fixed-up-projection} and \eqref{eq:fixed-gate-projection} element wise gives the rightmost vector of Eq.~\eqref{eq:logic-module-as-block-linear-map}.
Finally, choose the down-projection as $\mymat{W_d}:=\bigl(\mymat{M_1}, \dots, \mymat{M_{d_g}}\bigr)\in\RR^{N\times d_g N}$. 
With this choice, Eqs.~\eqref{eq:logic-module-as-block-linear-map}--\eqref{eq:fixed-gate-projection} imply $\myvec{z}_{\mathrm{MLP}}=\myvec{z}$ for the linear activation $\phi(x)=x$. 
Thus, the specialized logic module is representable as a GLU-style MLP with fixed up-projection and gate-projection weights and learnable down-projection blocks.

\paragraph{Embeddings}
Similarly to the logic module, the attention head can be implemented from a default attention architecture.
With the bundled embedding structure introduced above, i.e., a $(d_g + N)$-dimensional embedding with one-hot vectors for the actions and states, we define key and query matrices $W_K, W_Q \in \RR^{d_h \times (d_g + N)}$ as
\begin{align}
    W_K
    = \begin{pmatrix}
        (W_K)_1 & \cdots & (W_K)_1 \\
        \vdots & & \vdots \\
        (W_K)_{d_h} & \cdots & (W_K)_{d_h}
    \end{pmatrix}
    \ , \quad 
    W_Q 
    = \begin{pmatrix}
        (W_Q)_1 & \cdots & (W_Q)_1 \\
        \vdots & & \vdots \\
        (W_Q)_{d_h} & \cdots & (W_Q)_{d_h}
    \end{pmatrix}
    \ .
\end{align}
The first column is repeated $d_g + N$ times.
In that regard, the weight matrix reduces to one learnable vector.
Then, any embedded vector $\myvec{u_i} \in \RR^{d_g + N}$ is one-hot, because it is either an action or a state.
This yields
\begin{align}
    W_K \begin{pmatrix}
        \myvec{g} \\
        \myvec{0}_N
    \end{pmatrix}
    = \begin{pmatrix}
        (W_Q)_1 \\
        \vdots \\
        (W_Q)_{d_h}
    \end{pmatrix}
    \ , \qquad 
    W_K \begin{pmatrix}
        \myvec{0}_{d_g} \\
        \myvec{s}
    \end{pmatrix}
    = \begin{pmatrix}
        (W_Q)_1 \\
        \vdots \\
        (W_Q)_{d_h}
    \end{pmatrix}
    \ ,
\end{align}
and equally for $W_Q$, for any one-hot encoded action $\myvec{g}$ or state $\myvec{s}$.
This is completely equivalent to choosing a unit embedding and key/query vectors such that $\myvec{W_K} \myvec{u_i} = \myvec{W_K}$ with $\myvec{u_i} = 1 \in \RR^{1}$.
Consequently, we may write the keys and queries as vectors and omit the embedding in the attention head.

% % TODO:
% The value matrix

\section{Derivation of the mean correct/other logits}

\subsection{Mean correct logit}
The next correct logit $z_{i+1, k}$ following position $i$ in the $L$-sequence can be computed directly. Predictions are made only for the states following positions $i = L+1, \dotsc, 2L$; and for these prediction positions, we have
\begin{align}
    z_{i+1,k} = \myvec{p_{i+1}^{\star}} \cdot \myvec{z_{i+1}}
\end{align}
where $\myvec{p_{i+1}^{\star}}$ is the teacher ``logit'' vector, a one-hot vector with unity at entry $k$. This gives
\begin{align}
    z_{i+1, k} &= \sum_{a, b = 1}^{d_g} \myvec{e_{x_{i}}^{\top}} (\mymat{M_{a}^{\star}})^{\top} \mymat{M_{b}} \myvec{e_{x_{i}}} \left( \sum_{j \leq i} \pi_{ij} (v_{j})_{b} \right) (v_{i-L})_{a}
\end{align}
Here, $\myvec{e_{x_{i}}} \in \RR^N$ is a state token written as a one-hot encoded (OHE) vector. 
Similarly, the value vector $\myvec{v}_{j}$ from the transformer is an OHE vector for the actions, but zero for the states:
\begin{align}
    \myvec{v}_{j} := \begin{cases}
        \myvec{\hat{e}_{g_j}} \ , \quad &j = 1, \dotsc, L \\
        \myvec{0} \ , \quad &j \geq L + 1
    \end{cases}
\end{align}
with unit vectors in $\RR^{d_g}$. 
Each $g_j$, $j = 1, \dotsc, L$, specifies the action among the $d_g$ permutations in the vocabulary. 

We combine the facts that $i \geq L + 1$ here and that $\myvec{v}_j$ is identically zero for $j \geq L + 1$ to rewrite the inner sum:
% \begin{align}
%     z_{i+1, k} = \sum_{a, b = 1}^{d_g} \Bigl[ (M_{a}^{\star})^{\top} M_{b} \Bigr]_{x_i x_i} \left( \sum_{j = 1}^{L} \pi_{ij} (v_{j})_{b} \right) (v_{i-L})_{a} 
% \end{align}
\begin{align}
    z_{i+1, k} 
    &= \sum_{a, b = 1}^{d_g} \Bigl[ \mymat{(M_{a}^{\star})^{\top}} \mymat{M_{b}} \Bigr]_{x_i x_i} \biggl( \sum_{j = 1}^{L} \pi_{ij} \delta_{g_j, b} \delta_{g_{i-L}, a} \biggr)
    \\
    &= \sum_{a, b = 1}^{d_g} \Bigl[ \mymat{(M_{a}^{\star})^{\top}} \mymat{M_{b}} \Bigr]_{x_i x_i} \biggl( \pi_{i, i-L} \delta_{g_{i-L}, b} \delta_{g_{i-L}, a} + \sum_{\substack{j = 1 \\ j \neq i-L}}^{L} \pi_{ij} \delta_{g_j, b} \delta_{g_{i-L}, a} \biggr)
\end{align}
One model trains on a large number of training examples.
To account for the average behavior, we have to take an expectation value w.r.t. the current state $x_i$ and all the actions $g_{j}$, $j \in \{ 1, \dotsc, L \}$.
Here, all actions are needed, because the attention output depends on all those actions via the Values.
Because the current state can be obtained by the chain of permutations, the average w.r.t. $x_i = s_{i-L-1}$ is equivalent to an average over the initial state $s_0$, when simultaneously averaging over the actions. 
We assume a uniform probability distribution for all quantities. 
These two averages are then statistically independent.

\begin{figure}[!tbh]
    \centering
    \includegraphics[width=0.82\linewidth]{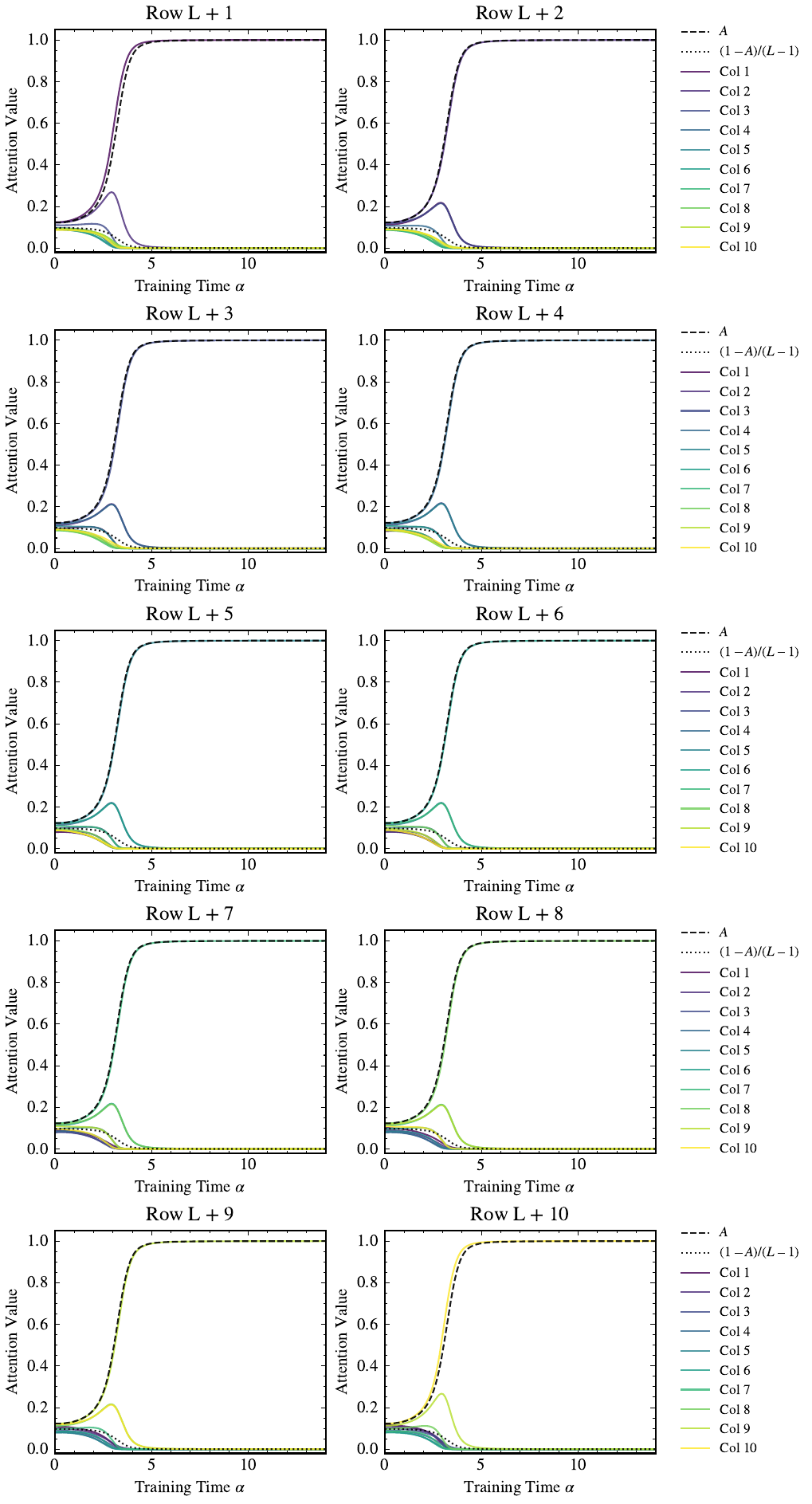}
    \caption{Entries of the attention block as a function of the learning time.}
    \label{fig:attention-entries}
\end{figure}

We apply the expectation value and rewrite $\delta_{g_{i-L}, b} \delta_{g_{i-L}, a} = \delta_{ab} \delta_{g_{i-L}, a}$:
\begin{align}
    \langle z_{i+1, k} \rangle
    &= \frac{1}{N} \sum_{x_i = 1}^{N} \frac{1}{d_g^L} \sum_{g_1 = 1}^{d_g} \sum_{g_2 = 1}^{d_g} \dotsc \sum_{g_L = 1}^{d_g} \\
    &\quad \times \Biggl( \sum_{a, b = 1}^{d_g} \Bigl[ \mymat{(M_{a}^{\star})^{\top}} \mymat{M_{b}} \Bigr]_{x_i x_i} \biggl( \pi_{i, i-L} \delta_{ab} \delta_{g_{i-L}, a} + \sum_{\substack{j = 1 \\ j \neq i-L}}^{L} \pi_{ij} \delta_{g_j, b} \delta_{g_{i-L}, a} \biggr) \Biggr)
\end{align}
Then, we get
\begin{align}
    \langle z_{i+1, k} \rangle 
    &= \sum_{a, b = 1}^{d_g} \frac{1}{N} \Tr \Bigl[ \mymat{(M_{a}^{\star})^{\top}} \mymat{M_{b}} \Bigr] \biggl( \pi_{i, i-L} \, \frac{\delta_{ab}}{d_g} + \sum_{\substack{j = 1 \\ j \neq i-L}}^{L} \pi_{ij} \frac{1}{d_g^2} \biggr) \\
    &= \sum_{a, b = 1}^{d_g} \frac{1}{N} \Tr \Bigl[ \mymat{(M_{a}^{\star})^{\top}} \mymat{M_{b}} \Bigr] \biggl[ \pi_{i, i-L} \biggl( \frac{\delta_{ab}}{d_g} - \frac{1}{d_g^2} \biggr) + \frac{1}{d_g^2} \biggr] \\
    &= \pi_{i, i-L} R + (1 - \pi_{i, i-L}) \frac{1}{d_g^2} \sum_{a, b = 1}^{d_g} \frac{1}{N} \Tr \Bigl[ \mymat{(M_{a}^{\star})^{\top}} \mymat{M_{b}} \Bigr] \\
    &= \biggl( \pi_{i, i-L} + \frac{1 - \pi_{i, i-L}}{d_g} \biggr) R + \frac{d_g - 1}{d_g} (1 - \pi_{i, i-L}) S
\end{align}
Let us assume that there is no preferred step in the sequence; then we can use the ansatz $\pi_{i+L, i} = A$.
This is empirically confirmed by the attention entries's evolution during training shown in Fig.\,\ref{fig:attention-entries}.
Consequently,
\begin{align}
    \langle z_{i+1, k} \rangle &= \biggl( A + \frac{1 - A}{d_g} \biggr) R + \frac{d_g - 1}{d_g} (1 - A) S
\end{align}
% This expression is independent of the sequence step $i$. 
% We can therefore define the mean correct logit, $\langle z_{i+1, k} \rangle := \langle z_{i+1, k} \rangle_{s_0, g}$, which is independent of $i \in \{ L+1, 2L \}$, 
% \begin{align}
%     \langle z_{i+1, k} \rangle &= \biggl( A + \frac{1 - A}{d_g} \biggr) \psi + \frac{d_g - 1}{d_g} \, (1 - A) S
% \end{align}

\begin{figure}[!tb]
    \centering
    \includegraphics[width=0.32\linewidth]{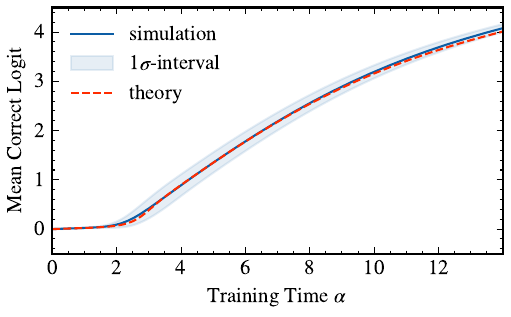}
    \includegraphics[width=0.32\linewidth]{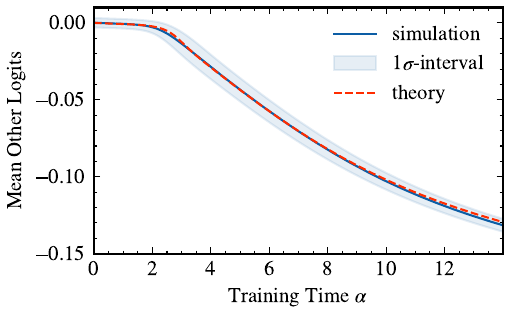}
    \includegraphics[width=0.32\linewidth]{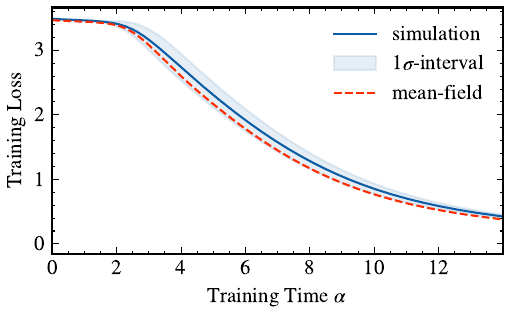}
    \caption{(a) Mean correct and other logits.
    (b) Mean-field loss ($\ell_{\mathrm{CE}}(A, R, S)$) compared to the actual loss}
    \label{fig:mean-correct-and-other-logits-and-loss}
\end{figure}

\subsection{Mean other logits}
Starting with the expression for the logit vector, we calculate the average logit first. For any $r = 1, \dotsc, N$, we have
\begin{align}
    z_{i+1, r} 
    &= \myvec{z}_{i+1} \cdot \myvec{e}_{r} = \sum_{a = 1}^{d_g} \myvec{e}_{r}^{\top} \mymat{M_{a}} \myvec{e}_{x_{i}} \sum_{j = 1}^{L} \pi_{ij} (v_j)_{a}
    = \sum_{a = 1}^{d_g} \bigl[ \mymat{M_{a}} \bigr]_{r x_i} \sum_{j = 1}^{L} \pi_{ij} \delta_{g_{j} a} 
\end{align}
and in the expectation value,  
\begin{align}
    \langle z_{i+1, r} \rangle &= \frac{1}{N} \sum_{x_i = 1}^{N} \frac{1}{d_g^L} \sum_{g_1 = 1}^{d_g} \sum_{g_2 = 1}^{d_g} \dotsc \sum_{g_L = 1}^{d_g} \sum_{a = 1}^{d_g} \bigl[ \mymat{M_{a}} \bigr]_{r x_i} \sum_{j = 1}^{L} \pi_{ij} \delta_{g_{j} a} \\
    &= \frac{1}{d_g N} \sum_{s = 1}^{N} \sum_{a = 1}^{d_g} \bigl[ \mymat{M_{a}} \bigr]_{r s} \sum_{j = 1}^{L} \pi_{ij}
\end{align}
The sum over the attention rows are exactly 1 for all $i \geq L+1$ with our construction. 
The sum of all logits remains constant throughout the training due to the combination of cross-entropy loss and gradient flow. 
% This is certainly the case for the exact logits and, by extension, for the mean logits as well because of the linear operation of the expectation value.
Hence, we define the constant parameter $\zeta_{\mathrm{init}}$ as the sum of the logits:
\begin{align}
    \zeta_{\mathrm{init}} &:= \frac{1}{d_g N} \sum_{a = 1}^{d_g} \sum_{r, s = 1}^{N} \bigl[ \mymat{M_{a}} \bigr]_{rs} 
\end{align}

Now, we obtained the average expression for \emph{any} logit. 
That is, for some initializations, $r$ is the correct logit's index; for many others, $r$ is the index for some other logit. 
As we are interested in the definite, non-correct logit here and already have an identity for the mean correct logit, we define
\begin{align}\label{eq:mean-wrong-logit}
    \langle z_{i+1, \overline{k}} \rangle := \frac{1}{N - 1} \Bigl( \zeta_{\mathrm{init}} - \langle z_{i+1, k} \rangle \Bigr)
\end{align}
As $\zeta_{\mathrm{init}}$ is fundamentally a quantity depending on the model initialization, different seeds will yield different dynamics of the mean other logits and thus of the order parameters.

\section{Differential equations for the order parameters}
The micro-parameter change, obtained from the discrete learning update step, is described by
\begin{align}
    \Delta [M_{a}]_{rs}(\alpha) := [M_{a}]_{rs}(\alpha + \Delta \alpha) - [M_{a}]_{rs}(\alpha) = - \eta \, \frac{\diff \ell_{\mathrm{CE}}}{\diff [M_{a}]_{rs}} 
\end{align}
We write the derivative w.r.t. the micro-parameters via the mean correct/other logits with the chain rule.
For them, we find the relation
\begin{align}
    \frac{\partial \ell_{\mathrm{CE}}}{\partial \langle z_{i+1, k} \rangle}
    = - \frac{\partial \ell_{\mathrm{CE}}}{\partial \langle z_{i+1, \overline{k}} \rangle}
\end{align}

\paragraph{Mean correct/other logit derivatives}
 Using the expression for the mean correct logit \eqref{eq:mean-correct-logit}, it is
\begin{align}
    \frac{\diff \langle z_{i+1, k} \rangle}{\diff [M_{b}]_{rs}} &= \frac{\partial \langle z_{i+1, k} \rangle}{\partial R} \frac{\diff R}{\diff [M_{b}]_{rs}} + \frac{\partial \langle z_{i+1, k} \rangle}{\partial S} \frac{\diff S}{\diff [M_{b}]_{rs}} 
    \\
    &= c_{R} \, \frac{\diff R}{\diff [M_{b}]_{rs}} + c_{S} \, \frac{\diff S}{\diff [M_{b}]_{rs}} 
\end{align}
There is no derivative of $A$ here, because $A$ does not depend on the learned permutation matrices at all.
The parameters $c_{R}, c_{S}$ are given by % $\alpha_0 \approx 3.20, \alpha_1 \approx 0.04$ and
\begin{align}
    c_{R} := A + \frac{1 - A}{d_g} \ , \quad c_S = \frac{d_g - 1}{d_g} (1 - A) = 1 - c_{R} % = 1 - A - \frac{1 - A}{d_g} 
\end{align}
One can show that
\begin{align}
    \frac{\diff R}{\diff [M_{b}]_{rs}} &= \frac{1}{d_g N} \, [M^{\star}_{b}]_{rs} \ , \qquad \frac{\diff S}{\diff [M_{b}]_{rs}} = \frac{1}{d_g (d_g - 1) N} \sum_{\substack{c = 1 \\ c \neq b}}^{d_g} [M^{\star}_{c}]_{rs}
\end{align}
Thus,
\begin{align}
    \frac{\diff \langle z_{i+1, k} \rangle}{\diff [M_{b}]_{rs}} &= \frac{c_{R}}{d_g N} \, [M^{\star}_{b}]_{rs} + \frac{c_{S}}{d_g (d_g - 1) N} \sum_{\substack{c = 1 \\ c \neq b}}^{d_g} [M^{\star}_{c}]_{rs} 
\end{align}

Take the expression for the mean other logit \eqref{eq:mean-wrong-logit}, then 
\begin{align}
    \frac{\diff \langle z_{i+1, \overline{k}} \rangle}{\diff [M_{b}]_{rs}} &= - \frac{1}{N - 1} \frac{\diff \langle z_{i+1, k} \rangle}{\diff [M_{b}]_{rs}} + \frac{1}{N - 1} \frac{1}{d_g N} 
\end{align}
Before we resolve this expression, it is best to advance to the update for $[M_{b}]_{rs}$.

\paragraph{The micro-parameter update.} Due to the opposing minus signs in the mean correct/other logits in the loss expression, we get a simplification in
\begin{align}
    \Delta [M_{b}]_{rs} &= - \eta \biggl( \frac{\partial \ell_{\mathrm{CE}}}{\partial \langle z_{i+1, k} \rangle} \frac{\diff \langle z_{i+1, k} \rangle}{\diff [M_{b}]_{rs}} + \frac{\partial \ell_{\mathrm{CE}}}{\partial \langle z_{i+1, \overline{k}} \rangle} \frac{\diff \langle z_{i+1, \overline{k}} \rangle}{\diff [M_{b}]_{rs}} \biggr) \\
    &= - \eta \, \frac{\partial \ell_{\mathrm{CE}}}{\partial \langle z_{i+1, k} \rangle} \frac{N}{N-1} \biggl( \frac{\diff \langle z_{i+1, k} \rangle}{\diff [M_{b}]_{rs}} - \frac{1}{d_g N^2} \biggr)
\end{align}
and we can now plug in the mean correct logit derivative:
\begin{align}
    \Delta [M_{b}]_{rs} &= - \frac{\eta}{d_{g} N} \frac{\partial \ell_{\mathrm{CE}}}{\partial \langle z_{i+1, k} \rangle} \frac{N}{N-1} \Biggl[ c_{R} \, [M^{\star}_{b}]_{rs} + \frac{c_{S}}{d_g - 1} \sum_{\substack{c = 1 \\ c \neq b}}^{d_g} [M^{\star}_{c}]_{rs} - \frac{1}{N} \Biggr]
\end{align}

\paragraph{Order parameter updates}
With the order parameter definitions of $R$ and $S$, we find
\begin{align}
    \Delta R 
    &= \frac{1}{d_g} \sum_{a = 1}^{d_g} \frac{1}{N} \Tr \Bigl[ (M^{\star}_{a} )^{\top} \Delta M_{a} \Bigr]
    \\
    &= - \frac{\eta}{d_g N} \, \frac{\partial \ell_{\mathrm{CE}}}{\partial \langle z_{i+1, k} \rangle} \frac{N}{N-1} \biggl( c_{R} + c_{S} \tau - \frac{1}{N} \biggr)
\end{align}
because the teacher matrices are fixed for one seed, and
\begin{align}
    \Delta S
    &= \frac{1}{d_g (d_g - 1)} \sum_{a = 1}^{d_g} \sum_{\substack{b = 1 \\ b \neq a}}^{d_g} \frac{1}{N} \Tr \Bigl[ (M^{\star}_{a} )^{\top} \Delta M_{b} \Bigr]
    \\
    &= - \frac{\eta}{d_g N} \, \frac{\partial \ell_{\mathrm{CE}}}{\partial \langle z_{i+1, k} \rangle} \frac{N}{N-1} \biggl( c_{R} \tau + c_{S} \, \frac{d_{g} - 2}{d_{g} - 1} \, \sigma + \frac{c_{S}}{d_{g} - 1} - \frac{1}{N} \biggr)
\end{align}
Two constants appear, $\tau$ and $\sigma$.
They are defined as
\begin{align}
    \tau 
    &:= \frac{1}{d_g (d_g - 1) N} \sum_{a = 1}^{d_g} \sum_{\substack{c = 1 \\ c \neq a}}^{d_g} \Tr \Bigl[ \bigl( \mymat{M^{\star}_{a}} \bigr)^{\top} \mymat{M^{\star}_{c}} \Bigr]
\end{align}
and
\begin{align}
    \sigma 
    &:= \frac{1}{d_g (d_g - 1) (d_g - 2) N} \sum_{a = 1}^{d_g} \sum_{\substack{b = 1 \\ b \neq a}}^{d_g} \sum_{\substack{c = 1 \\ c \neq a, b}}^{d_g} \Tr \Bigl[ \bigl( M^{\star}_{a} \bigr)^{\top} M^{\star}_{c} \Bigr]
\end{align}
and depend on the seed.
We readily find $\sigma = \tau$ and are left with one such permutation matrix trace.
As permutation matrices are not orthogonal, there is some overlap.

Many seeds give rise to many choices for the set of permutations.
We can average over the seeds to get $\overline{\tau} = 1 / N + \varepsilon$ with a correction $\varepsilon = - (1-1/N) / (N! - 1)$, which is negligible for non-small $N$.
Details on this average are given in Appendix \ref{app:permutation-trace}.
The final difference quotients using our mean-field cross-entropy then read
\begin{subequations}
    \begin{align}
        \Delta R &= \frac{\eta}{d_g N} \, \frac{1}{1 + \frac{1}{N - 1} \, \ee^{m(A, R, S)}} \biggl( A + \frac{1 - A}{d_g} \biggr) 
        \\
        \Delta S &= \frac{\eta}{d_g N} \, \frac{1}{1 + \frac{1}{N - 1} \, \ee^{m(A, R, S)}} \frac{1 - A}{d_g}  
    \end{align}
\end{subequations}
with $m(A, R, S) = \frac{1}{L} \sum_{i = L+1}^{2L} \langle z_{i+1, k} \rangle - \langle z_{i+1, \overline{k}} \rangle$

\paragraph{The attention order parameter} 
To see the dynamics of $A$, we write, for the moment, $\myvec{W} = (\myvec{W_{K}^{\top}}, \myvec{W_{Q}^{\top}})^{\top}$ as a vector of all key and query weights.
We calculate
\begin{align}
    d_g N (\nabla_{\myvec{W}} A) \cdot (- \eta) (\nabla_{\myvec{W}} \ell_{\mathrm{CE}})
    &= \nabla_{\myvec{W}} A \cdot \frac{\Delta \myvec{W}}{\Delta \alpha} 
    \to \nabla_{\myvec{W}} A \cdot \frac{\diff \myvec{W}}{\diff \alpha}
    = \frac{\diff A}{\diff \alpha} 
\end{align} 
Applying the chain rule once more for $\nabla_{\myvec{W}} \ell_{\mathrm{CE}} = (\nabla_{\myvec{W}} A) \frac{\diff \ell_{\mathrm{CE}}}{\diff A}$ (only $A$ depends on the weights) gives 
\begin{align}
    \frac{\diff A}{\diff \alpha} &= - \eta d_g N (\nabla_{\myvec{W}} A)^2 \frac{\diff \ell_{\mathrm{CE}}}{\diff A}
\end{align}
when assuming the right-hand side is finite in the thermodynamic limit.
It will be seen that this is indeed the case.

Define attention logits $l_{ij} = \myvec{q_i^\top} \myvec{k_j} / \sqrt{d_h} = \myvec{W_{Q}^{\top}} \mymat{\Omega(j - i)} \myvec{W_{K}} / \sqrt{d_h}$.
Then the gradient $\nabla_{\myvec{W}} A$ becomes
\begin{align}
    \nabla_{\myvec{W}} A &= \sum_{i = L+1}^{2L} \frac{\partial A}{\partial \pi_{i, i-L}} \, \nabla_{\myvec{W}} \pi_{i, i-L} = \sum_{i = L+1}^{2L} \frac{\partial A}{\partial \pi_{i, i-L}} \sum_{k = 1}^{L} \frac{\partial \pi_{i, i-L}}{\partial l_{ik}} \, \nabla_{\myvec{W}} l_{ik} 
    % &= \sum_{i = L+1}^{2L} \sum_{k = 1}^{L} \frac{\partial A}{\partial \pi_{i, i-L}} \frac{\partial \pi_{i, i-L}}{\partial l_{ik}} \sum_{\mu = 1}^{d_h} \biggl( \frac{\partial l_{ik}}{\partial [W_{Q}]_{\mu}} \frac{\Delta [W_{Q}]_{\mu}}{\Delta \alpha} + \frac{\partial l_{ik}}{\partial [W_{K}]_{\mu}} \frac{\Delta [W_{K}]_{\mu}}{\Delta \alpha} \biggr)
\end{align}
Let us define the function
\begin{align}
    c(A) 
    &:= d_g N (\nabla_{\myvec{W}} A)^2 
    = d_g N \Biggl[ \sum_{i, k = 1}^{L} \frac{\partial A}{\partial \pi_{i+L, i}} \frac{\partial \pi_{i+L, i}}{\partial l_{i+L, k}} \Bigl( \nabla_{\myvec{W}} l_{i+L, k} \Bigr) \Biggr]^2 
\end{align}
then the differential equation reads
\begin{align}
     \frac{\diff A}{\diff \alpha} 
     &= \eta \, c(A) \, \frac{1}{1 + \frac{1}{N - 1} \, \ee^{m(A, R, S)}} \frac{N}{N - 1} \frac{d_g - 1}{d_g} (R - S) 
\end{align}
so that merely $c(A)$ needs to be yet determined. 
Given the derivatives
\begin{align}
    \frac{\partial A}{\partial \pi_{i+L, i}} &= \frac{1}{L}
    \\
    \frac{\partial \pi_{i+L, i}}{\partial l_{i+L, k}} &= \pi_{i+L, i} (\delta_{ik} - \pi_{i+L, k}) % =: a_{ik}
    \\
    \nabla_{\myvec{W_Q}} l_{i+L,k} &= \frac{\mymat{\Omega(k - i - L)} \myvec{W_K}}{\sqrt{d_h}}
    \\
    \nabla_{\myvec{W_K}} l_{i+L,k} &= \frac{\mymat{\Omega(i + L - k)} \myvec{W_Q}}{\sqrt{d_h}}
\end{align}
some linear algebra yields, for the approximation $\pi_{i+L, j} = A \delta_{ij} + \frac{1 - A}{L - 1} (1 - \delta_{ij})$ of an expected seed-averaged behavior,
\begin{align}
    c(A) 
    &= d_g N A^2 (1 - A)^2 \frac{\myvec{W_K^{\top}} (\Id - \mymat{\Lambda})^2 \myvec{W_K} + \myvec{W_Q^{\top}} (\Id - \mymat{\Lambda})^2 \myvec{W_Q}}{d_h}
\end{align}
where $\mymat{\Lambda} = \mathrm{diag} (\lambda_{1}, \lambda_{1}, \lambda_{2}, \lambda_{2}, \dotsc, \lambda_{d_h/2}, \lambda_{d_h/2})$ with $\lambda_{n} = 1$ for $\omega_{n} \in 2 \pi \mathbb{Z}$ or otherwise
\begin{align}
    \lambda_{n} &= \frac{1}{L - 1} \biggl[ \frac{1}{L} \biggl( \frac{\sin(\omega_{n} L / 2)}{\sin(\omega_{n} / 2)} \biggr)^2 - 1 \biggr]
\end{align}
While the diagonal part in $\pi_{i+L, j} = A \delta_{ij} + \frac{1 - A}{L - 1} (1 - \delta_{ij})$ approximates the attention entries well, the off-diagonal entries behave differently during training.
This elongates the smooth increase of $A$ in the learning regime.
Without the uneven behavior, the slope of $A$ would grow too steeply, because $A^2 (1 - A)^2$ acts as a catalyst in the differential equation for $A$. 
At the end of training, just as at the very beginning, this approximation describes the system well, however.
Hence, the above expression for $c(A)$ may be used to determine the value at initialization, where the ansatz is still correct.
We find that the empirical trajectory is approximated by $c(A) = Lc_{\omega} \cdot A (1 - A)$ with
\begin{align}
    c_{\omega} &:= \frac{d_g N}{L} \, \frac{L - 1}{L^2} \Biggl\langle \frac{\myvec{W_K^{\top}} (\Id - \mymat{\Lambda})^2 \myvec{W_K} + \myvec{W_Q^{\top}} (\Id - \mymat{\Lambda})^2 \myvec{W_Q}}{d_h} \Bigg\vert_{\alpha = 0} \Biggr\rangle_{\myvec{W_Q}, \myvec{W_K}} 
\end{align}
At initialization, $\myvec{W_Q}$ and $\myvec{W_K}$ are random vectors, such that, for some random $\myvec{X} \in \RR^{d_h}$,
\begin{align}
    \langle \myvec{X^{\top}} (\Id - \mymat{\Lambda})^2 \myvec{X} \rangle_{\myvec{X}}
    &= \Tr \langle \myvec{X^{\top}} (\Id - \mymat{\Lambda})^2 \myvec{X} \rangle_{\myvec{X}}
    = \Tr \Bigl[ (\Id - \mymat{\Lambda})^2 \langle \myvec{X^{\top}} \myvec{X} \rangle_{\myvec{X}} \Bigr]
\end{align}
exploiting the linearity of the trace and expectation value.
$\langle \myvec{X} \myvec{X^{\top}} \rangle_{\myvec{X}}$ is the covariance matrix of $\myvec{X}$. 
An i.i.d. initialization yields a scaled unit matrix with, say, $\sigma^2$ as the variance.
Then $\langle \myvec{X^{\top}} (\Id - \mymat{\Lambda})^2 \myvec{X} \rangle_{\myvec{X}} = \sigma^2 \Tr \bigl[ (\Id - \mymat{\Lambda})^2 \bigr]$ and
\begin{align}
    c_{\omega} 
    &= \frac{d_g N}{L} \frac{L - 1}{L^2} \, \sigma^2 \frac{2}{d_h} \Tr \bigl[ (\Id - \mymat{\Lambda})^2 \bigr] 
\end{align}
Plugging in the definition of $\mymat{\Lambda}$ followed by some algebraic manipulations leads to the final expression in Eq.\,\eqref{eq:c-omega-prefactor}.

In principle, we can determine a discrete update step $\Delta A$ via a Taylor expansion in the $\Delta l_{ij}$ of the softmax activation.
These, in turn, are expressible via $\Delta \myvec{W_Q}, \Delta \myvec{W_K}$ resulting in a series in orders of the learning rate. 
The gradient-flow limit $\eta \to 0$ lets these higher-order terms vanish.
Small learning rates allow us to neglect them, yielding a first-order expression equivalent to the above derivative.

\section{Seed-averaged permutation matrix overlap}
\label{app:permutation-trace}
Our goal is to find the quantity $\Tr \bigl[ (\mymat{M^{\star}_{a})^{\top}} \mymat{M^{\star}_{b}} \bigr]$ for some $a \neq b$, $a, b \in \{ 1, \dotsc, d_g \}$ on average. 
Because a product of two permutation matrices is again a permutation matrix, we reduce the problem of finding the overlap of two random matrices by computing the average trace of a permutation. 
Thus, w.l.o.g., we can assume one of them to be the unit matrix. Since the trace is a linear operator, we can compute the average permutation matrix itself. 
As each entry in the first row has a probability of $1/N$ to have a non-zero entry when constructing such a matrix, and because each row must behave alike due to symmetry, the average permutation matrix is filled with the entries $1/N$.

Now, we only want to consider overlaps of different permutation matrices, thus we have to consider only non-trivial matrices when assuming the other one is the identity. 
To account for that, we write 
\begin{align}
    \langle \mymat{M^{\star}} \rangle_{\mymat{M^{\star}} \in S^{N} \setminus \{ \Id_N \}} &= \frac{1}{N! - 1} \Biggl( \sum_{i = 1}^{N!} \mymat{M_i^{\star}} - \Id_N \Biggr) = \frac{1}{N! - 1} \biggl( \frac{N!}{N} J_{N} - \Id_N \biggr)
\end{align}
with $(J_N)_{rs} = 1$ $\forall r, s = 1, \dotsc, N$ and $N \geq 2$. 
Hence,
\begin{align}
    \langle \Tr \mymat{M}^{\star} \rangle_{\mymat{M}^{\star} \in S^{N} \setminus \{ \Id_N \}} &= \frac{N!}{N! - 1} - \frac{N}{N! - 1} = \frac{1 - \frac{1}{(N-1)!}}{1 - \frac{1}{N!}}
\end{align}
With this intermediate finding, we may write 
\begin{align}
    \Bigl\langle \Tr \Bigl[ (\mymat{M}^{\star}_{a})^{\top} \mymat{M}^{\star}_{b} \Bigr] \Bigr\rangle_{\mymat{M}^{\star}_{a} \neq \mymat{M}^{\star}_{b} \in S^{N}} &= \frac{1 - \frac{1}{(N-1)!}}{1 - \frac{1}{N!}}
\end{align}
when $a \neq b$. 
We immediately realize that this constant approaches $1$ for $N \to + \infty$. 
Therefore, 
\begin{align}
    \overline{\tau} 
    &:= \langle \tau \rangle_{\mymat{M}^{\star}_{a} \neq \mymat{M}^{\star}_{b} \in S^{N}}
    = \frac{1}{N} \frac{1 - \frac{1}{(N-1)!}}{1 - \frac{1}{N!}}
    = \frac{1}{N} - \frac{1}{N} \frac{N - 1}{N! - 1} 
\end{align}

\section{Early training regime} \label{App:EarlyTrainingPhase}
At the start of training, $A$ remains on a plateau of $A = A_0$, which is $1/L$ on average (every attention entry is equal with random key/query initializations).
A Taylor expansion of the loss up to linear order describes the system well during the early training process.
The differential equations are solvable exactly in this regime:
\begin{align}
    A(\alpha) = A_0 \ , \qquad R(\alpha) = \eta \, \frac{N - 1}{N} \biggl( A_0 + \frac{1 - A_0}{d_g} \biggr) \cdot \alpha \ , \qquad S(\alpha) = \eta \, \frac{N - 1}{N} \frac{1 - A_0}{d_g} \cdot \alpha
\end{align}
The matrix overlaps $R$ and $S$ increase linearly before higher-order terms in the loss become relevant and the attention mechanism starts to learn.
We observe that the correct overlap increases faster than the overlap with wrong permutations.

\section{Onset of attention learning}
In the main text, we observed initially a flat attention order parameter $A$, followed by a rapid increase at the start of the second learning phase (see left panel of Fig.~\ref{fig:dynamics-constant-prefactor-approx}). 
Using this separation of learning into the two phases, we can analytically obtain an approximate solution of the differential equations for the order parameters that describe the onset of attention learning. 
Combining Eqs.~\eqref{Eq:diffeq_R} and \eqref{Eq:diffeq_S}, in the linearized regime of the first phase of training (Appendix \ref{App:EarlyTrainingPhase}) where $A(0) = 1/L$, we find
\begin{align}
    \frac{{\rm d} (R-S)}{{\rm d}\alpha} &\approx \eta \, \frac{N-1}{N} \frac{1}{L}\ .
\end{align}
In the limit of large $N$, the solution is thus given by $R-S \approx \alpha / L$. 
Using this in the differential equation for the attention order parameter, Eq.~\eqref{Eq:diffeq_A}, yields, to leading order in $1/L$,
\begin{align}
    \frac{{\rm d}A}{{\rm d}\alpha} &\approx \eta^2 c_\omega \alpha A  \\ 
    \Rightarrow A(\alpha) &\approx \frac{1}{L} \exp\left[\frac{c_\omega}{2} \eta^2 \alpha^2\right] \ .
\end{align}
This approximation is valid until the onset of the second learning phase.
The timescale of the increase is set by $\sqrt{2/(c_\omega \eta)} = O(1)$. % missing an \alpha^2 here?

\section{Final rollout accuracy modeling}
\label{app:rollout-accuracy-derivation}
To model the final rollout accuracy, we start with the recursion relation
\begin{align}
    P(\hat{s}_{t+1} = s_{t+1}) 
    &= P(\hat{s}_{t} = s_{t}) \cdot P(\hat{s}_{t+1} = s_{t+1} | \hat{s}_{t} = s_{t}) 
    \nonumber \\
    &\quad + \bigl[ 1 - P(\hat{s}_{t} = s_{t}) \bigr] \cdot P(\hat{s}_{t+1} = s_{t+1} | \hat{s}_{t} \neq s_{t})
\end{align}
with $P(\hat{s}_{0} = s_{0}) = 1$.
We shorten the notation:
\begin{align}
    \varrho_{t} := P(\hat{s}_{t} = s_{t}) \qquad 
    \mathcal{A}_{t} := P(\hat{s}_{t+1} = s_{t+1} | \hat{s}_{t} = s_{t}) \qquad
    \mathcal{B}_{t} := P(\hat{s}_{t+1} = s_{t+1} | \hat{s}_{t} \neq s_{t})
\end{align}
Then
\begin{align}
    \varrho_{t+1} &= \varrho_{t} (\mathcal{A}_{t} - \mathcal{B}_{t}) + \mathcal{B}_{t}
\end{align}
with $\varrho_{0} = 1$ for the initial token, and we seek $\varrho_{L}$ from this recursion relation. 
It holds
\begin{align}
    \varrho_{L}
    &= \mathcal{A}_{0} \prod_{t = 1}^{L-1} (\mathcal{A}_{t} - \mathcal{B}_{t}) + \sum_{t = 1}^{L-1} \Biggl[ \mathcal{B}_{t} \prod_{i = t+1}^{L-1} (\mathcal{A}_{i} - \mathcal{B}_{i}) \Biggr]
\end{align}

The next state is chosen by the maximal argument of the logit vector. 
That is, $\mathcal{A}_{t}$ is the probability of choosing the correct logit (when operating on $s_{t}$). 
From the definition of the logit, we see that the logit is the $s_{t}$-th column of our action-conditioned map, which does not depend on the last state token $s_{t-1}$.
If we assume that the other logits are equally likely on average, and that the prediction of state $i$ behaves the same as that of state $j$, then we may write
%and that the prediction behaves the same for each state token, i.e., each column in the map has the same distribution with just the correct logit in a different position, then we may write 
\begin{align}
    \mathcal{B}_{t} 
    &\approx \frac{1 - \mathcal{A}_{t}}{N - 1}
\end{align}
% $\approx$ here because: with a given wrong state, another column has another probability on the "correct" logit ("correct" for a different state)
% this approx. is more like the average over all N-1 columns (of the wrong states)
Next, we need to find an expression for the probability of choosing the correct logit (given the current state is correct). 
For this, we model the correct logit as a Gaussian variable.
A numerical check quickly supports this assumption: the correct logit is for different training examples and for each sequence step almost exactly distributed normally. % \red{Reference a figure!}

Viewing every sequence-step as the same, we can replace $\mathcal{A}_{t}$ by $\overline{\mathcal{A}} := \mathbb{E}_{t \in [L]} [\mathcal{A}_{t}]$ such that
\begin{align}
    \varrho_{L} 
    &\approx \frac{N - 1}{N} \biggl( \frac{N \overline{\mathcal{A}} - 1}{N - 1} \biggr)^{L} + \frac{1}{N}
\end{align}
after some rearrangements and exploitation of a geometric sum with $\sum_{n = 0}^{L-2} r^{n} = (1 - r^{L-1}) / (1 - r)$ and $r = \frac{N \overline{\mathcal{A}} - 1}{N - 1}$.
Here, the singularity is hit when $\overline{\mathcal{A}} = 1$, which cannot occur as the student is never able to achieve perfect predictions. 

Immediately, we realize that $\varrho_{L} \to 1$ when $\overline{\mathcal{A}} \to 1$. 
Random guessing ($\overline{\mathcal{A}} = 1/N$) would yield $\varrho_{L} = 1/N$ -- a random choice.
Yet $\varrho_{L}$ grows much slower: each token in the sequence has to be predicted correctly in a row.

The probability of the correct choice given the correct current logit can be expressed as
\begin{align}
    \overline{\mathcal{A}} = P(z_{k} > \max_{f \neq k} z_{f}).
\end{align}
This is equivalent to determining the probability of every other logit being at most $x$ (a certain correct logit $z_k$) and then integrating over $x$ including the Gaussian PDF of $x = z_{k}$:
\begin{align}
    \overline{\mathcal{A}} 
    &= \int_{- \infty}^{+ \infty} \diff x \, \frac{\ee^{- (x - \mu_k)^2 / (2 \sigma_k^2)}}{\sqrt{2 \pi} \sigma_k} \, \Phi \! \biggl( \frac{x - \mu_{\overline{k}}}{\sigma_{\overline{k}}} \biggr)^{N-1}
\end{align}
where $\Phi$ is the CDF of the Gaussian PDF:
\begin{align}
    \Phi(u) := \int_{- \infty}^{u} \diff y \, \frac{\ee^{- y^2 / 2}}{\sqrt{2 \pi}}
\end{align}
Upon substituting $u = (x - \mu_k) / \sigma_k$: 
\begin{align}
    \overline{\mathcal{A}}
    &= \int_{-\infty}^{+\infty} \diff u \, \frac{\ee^{- u^2 / 2}}{\sqrt{2 \pi}} \, \Phi \! \biggl( \frac{\sigma_{k}}{\sigma_{\overline{k}}} \, u + \frac{\mu_{k} - \mu_{\overline{k}}}{\sigma_{\overline{k}}} \biggr)^{N-1},
\end{align}
% Numerical integrations: use $\mu_k$, $\mu_{\overline{k}}$, $\sigma_k$, $\sigma_{\overline{k}}$ from model runs, averaged over seeds. 

% \begin{figure}
%     \centering
%     \includegraphics[width=0.48\linewidth]{neurips_plots/003_logitvariances.pdf}
%     \includegraphics[width=0.48\linewidth]{neurips_plots/004_rollout_accuracy_halfway_variance.pdf}
%     \caption{(a) Variances for the mean correct and other logits during training.
%     (b) Simulated rollout accuracy for the half-value between initialization and maximum: $(\max_{\alpha} (\sigma_{k}^2) - \sigma_{k}^2 \vert_{\alpha = 0}) / 2$}
%     \label{fig:logit-variances-and-accuracy-with-halway-variances}
% \end{figure}
\begin{figure}[!h]
    \centering
    \includegraphics[width=0.48\linewidth]{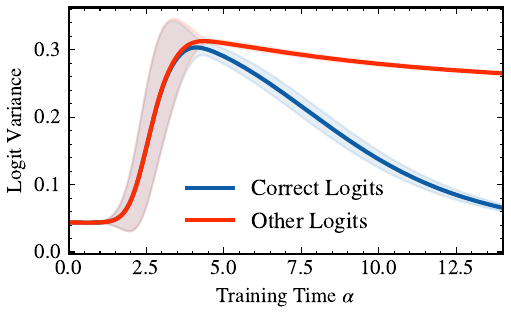}
    \includegraphics[width=0.48\linewidth]{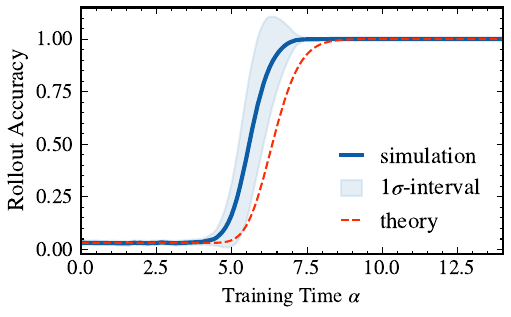}
    \caption{(a) Variances for the mean correct and other logits during training.
    (b) Simulated rollout accuracy for the empirical logit means and variances, where the empirical and theoretically predicted logit means had a near-perfect agreement.}
    \label{fig:logit-variances-and-accuracy-with-empirical-variances}
\end{figure}

\newpage
\section{Robustness concerning hyperparameters}
\label{app:robustness-hyperparameters}

\begin{table}[h]
  \caption{Different set of hyperparameters for training simulations.}
  \label{tab:config-old}
  \centering
  \begin{tabular}{lll}
    \toprule
    Variable & Value & Description \\
    \midrule
    $N$ & $24$ & Magnitude of permutation set $\{1, \dots, N\}$ \\
    $L$ & $10$ & Sequence steps (actions $g_1, \dots, g_L$; predicted states $s_1, \dots, s_L$) \\
    $d_g$ & $24$ & Vocabulary size of permutations \\
    $d_{\text{model}}$ & $1$ & Embedding dimension \\
    $d_h$ & $8$ & RoPE encoding dimension (key \& query dimension) \\
    $n_{\text{train\_seq}}$ & $32{,}768$ & Number of training examples \\
    $n_{\text{test\_seq}}$ & $128$ & Number of test examples \\
    $\text{batch\_size}$ & $32{,}768$ & Batch size for gradient descent (full-batch if equal to $n_{\text{train\_seq}}$) \\
    $\text{lr}$ & $5 \times 10^{-1}$ & Learning rate \\
    $\text{weight\_decay}$ & $0$ & Weight decay (Adam optimizer) \\
    $\text{rope\_theta}$ & $10{,}000$ & RoPE frequency base (power-law distributed) \\
    $\text{train\_epochs}$ & $16{,}000$ & Number of training epochs \\
    $\text{eval\_every}$ & $50$ & Evaluation frequency (in epochs) \\
    $\text{seeds}$ & $100$ & Number of seeds (one model per seed) \\
    $\text{optimizer}$ & \texttt{sgd} & Optimizer type (\texttt{sgd} or \texttt{adamw}) \\
    % $\text{logit\_scale}$ & \texttt{False} & Learnable logit scaling parameter $a_u$ \\
    $\text{fix\_embedding}$ & \texttt{True} & Restrict embedding vectors to unity \\
    % $\text{with\_softmax}$ & \texttt{True} & Use of softmax in attention head output \\
    \bottomrule
  \end{tabular}
\end{table}
Table \ref{tab:config-old} lists a different set of hyperparameters. 
The RoPE frequencies are here set as $\omega_{n} :=  \theta_{\mathrm{RoPE}}^{- 2 (n-1) / d_h}$ with $\theta_{\mathrm{RoPE}}$ (\verb|rope_theta|) as the base and $n = 1, \dots, d_h / 2$.

\begin{figure}[!tbh]
    \centering
    \includegraphics[width=\linewidth]{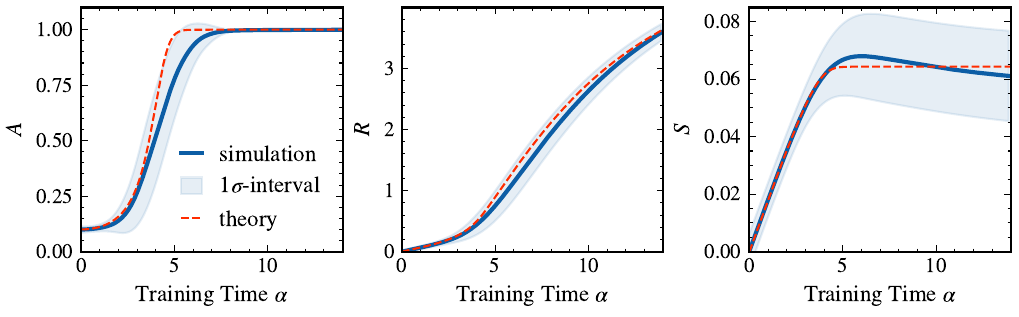}
    \caption{Order parameters from theory and simulations.}
    \label{fig:old-params_order-params}
\end{figure}

\begin{figure}[!tbh]
    \centering
    \includegraphics[width=0.48\linewidth]{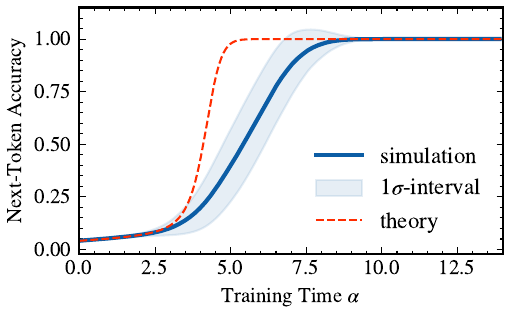}
    \includegraphics[width=0.48\linewidth]{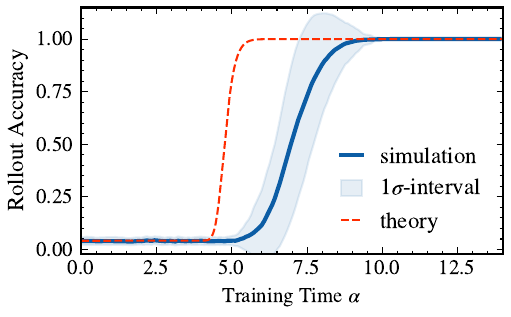}
    \caption{Next-token accuracy and final rollout accuracy for logit variances held fixed at their initial values of $\sigma^2_k = 0.0466$, $\sigma^2_{\overline{k}} = 0.0467$.}
    \label{fig:old-params_accuracies-for-initial-variances}
\end{figure}

\begin{figure}[!tbh]
    \centering
    \includegraphics[width=0.325\linewidth]{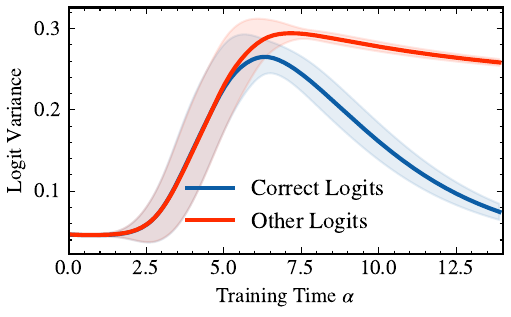}
    \includegraphics[width=0.325\linewidth]{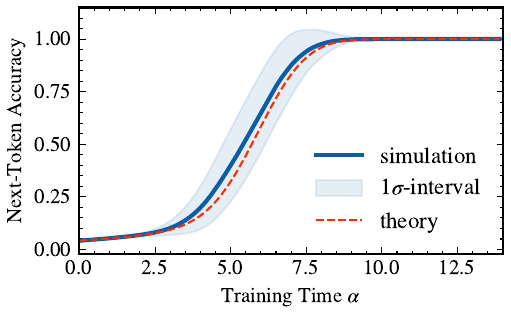}
    \includegraphics[width=0.325\linewidth]{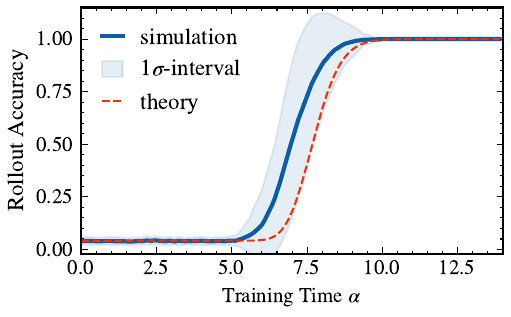}
    \caption{Next-token accuracy and final rollout accuracy with the empirical logit variances.}
    \label{fig:old-params_accuracies-for-empirical-variances}
\end{figure}

\begin{figure}[!tbh]
    \centering
    \includegraphics[width=0.325\linewidth]{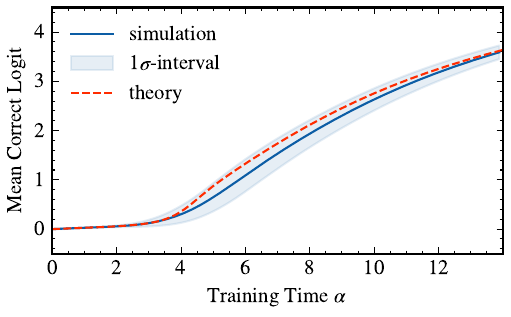}
    \includegraphics[width=0.325\linewidth]{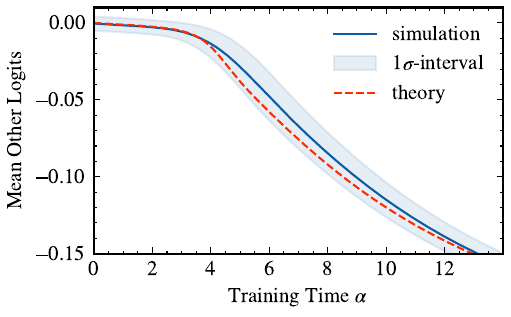}
    \includegraphics[width=0.325\linewidth]{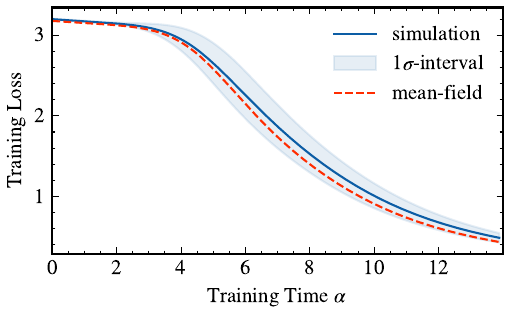}
    \caption{Mean correct/other logits and loss function.}
    \label{fig:old-params_logits-and-loss}
\end{figure}

The results are shown in Figures \ref{fig:old-params_order-params}, \ref{fig:old-params_accuracies-for-initial-variances}, \ref{fig:old-params_accuracies-for-empirical-variances}, and \ref{fig:old-params_logits-and-loss}.

\vspace{2cm}

\newpage 

\vspace{1cm}

\vspace{1cm}

% ======================================================================

% \newpage
% \ 
% \newpage 
% \ 
% \newpage
% \input{neurips_checklist.tex}

\end{document}